\documentclass[%
preprint,
preprintnumbers,
nofootinbib,
 amsmath,amssymb,
 aps,
 prd,
floatfix,
]{revtex4}

\usepackage{mathtools}
\usepackage{epsfig}
\usepackage{dcolumn}
\usepackage{bm}
\usepackage{amsthm}
\usepackage{amsmath}
\usepackage{amsfonts}
\usepackage{amssymb}
\usepackage{graphicx}
\usepackage{latexsym}
\usepackage{rotating}
\usepackage{multirow}
\usepackage{slashed}
\usepackage{color}
\usepackage{hyperref}
\usepackage{graphicx}%
\usepackage{subfigure}
\usepackage{booktabs}
\usepackage{diagbox}
\newcommand{\pt}{p_{\rm T}}

\hypersetup{
    colorlinks=true,
    linkcolor=blue,
    filecolor=magenta,      
    urlcolor=cyan,
}

\begin{document}

\title{Variational Autoencoders for Anomalous Jet Tagging}

\author{Taoli Cheng}
%\email{chengtaoli.1990@gmail.com}
\affiliation{Mila -- Quebec Artificial Intelligence Institute \\ 6666 St-Urbain, \#200, Montreal, QC, H2S 3H1, Canada}
\affiliation{Department of Informatics, Universit\'{e} de Montr\'{e}al}

\author{Jean-Fran\c{c}ois Arguin, Julien Leissner-Martin, Jacinthe Pilette}
\affiliation{D\'epartement de physique,
Universit\'e de Montr\'eal \\
Complexe des Sciences,
Case postale 6128, succursale Centre-ville,
Montr\'eal (Qu\'ebec)  H3C 3J7,
Canada}

\author{Tobias Golling}
\affiliation{Department of Particle Physics, University of Geneva \\
24, quai Ernest-Ansermet, CH-1211 Gen\`{e}ve 4, Switzerland }

\begin{abstract}
We present a detailed study on Variational Autoencoders (VAEs) for anomalous jet tagging at the Large Hadron Collider. By taking in low-level jet constituents' information, and training with background QCD jets in an unsupervised manner, the VAE is able to encode important information for reconstructing jets, while learning an expressive posterior distribution in the latent space. 
%The encoder (inference) and decoder (generation) can be used together or separately to identify out-of-distribution anomalous jets. 
%We employed different techniques to regularize the latent representation, and show how the behavior changes.
When using the VAE as an anomaly detector, we present different approaches to detect anomalies: directly comparing in the input space or, instead, working in the latent space.
%A comprehensive series of test sets are generated to fully examine the anomaly tagging performance in different jet types.
In order to facilitate general search approaches such as bump-hunt, mass-decorrelated VAEs based on distance correlation regularization are also studied. We find that the naive mass-decorrelated VAEs fail at maintaining proper detection performance, by assigning higher probability to some anomalous samples.
To build a performant mass-decorrelated anomalous jet tagger, we propose the Outlier Exposed VAE (OE-VAE), for which some outlier samples are introduced in the training process to guide the learned information.
OE-VAEs are employed to achieve two goals at the same time: increasing sensitivity of outlier detection and decorrelating jet mass from the anomaly score. We succeed in reaching excellent results from both aspects.
Code implementation of this work can be found at \href{https://github.com/taolicheng/VAE-Jet}{Github}.

\end{abstract}

\maketitle

%%%%%%%%%%%%%%%%%%%%%%%%%%%%%%%%%%%%%%%%%%%%%%%%%%%%%%%%%%%%%%%%%%%%%%
\clearpage

%\tableofcontents

\section{Introduction}

% background
Supervised classifiers based on deep neural networks have been used for boosted jet tagging and event selection at the LHC. They have become a mature research topic in the past few years. Despite the success of supervised taggers, the problem of searching for new physics signals suffer from unclear search channels and unpredictable signal properties.
It's thus worth exploring other methods beyond the specific supervised taggers to probe new physics signals in a model-independent manner. 
%Only null results for model-dependent new physics searches have been obtained so far at the LHC. 
Model-independent and data-driven approaches, especially assisted by modern machine learning techniques, are becoming potential alternative search strategies nowadays. Unsupervised learning methods including clustering, density estimation, etc. have been used in the general scope of detecting novel or anomalous events \cite{10.1145/1541880.1541882, Hodge2004ASO, PIMENTEL2014215}.
In applications for LHC physics, some anomaly detection methods \cite{Collins:2018epr, Collins:2019jip, Hajer:2018kqm, Nachman:2020lpy, Aguilar-Saavedra:2017rzt, Andreassen:2020nkr, Aad:2020cws, Dillon:2019cqt, Dillon:2020quc, Amram:2020ykb, Romao:2020ocr} including density estimation, weakly-supervised classification, etc. have been studied recently. 
%General anomaly detection methods: distance-based (fro high-dimensional data, it's hard to find a appropriate space in which the distance can be easily constructed for detecting anomalies.)
%One example is CWOLa \cite{Collins:2018epr, Collins:2019jip}, which makes use of the Neyman-Pearson lemma and extend classical bump hunt with...

\begin{figure}[htb!]
    \centering
    \includegraphics[width=0.8\textwidth]{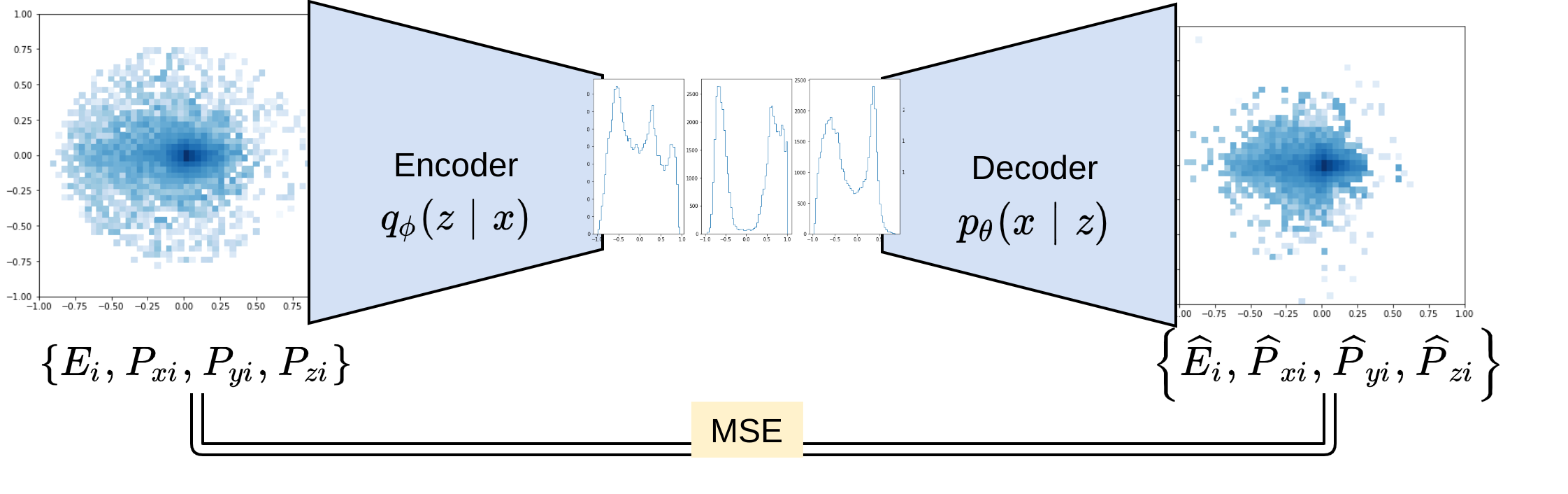}
    \caption{Schematic of Autoencoder-based anomalous jet tagger. Jet reconstruction is encouraged by minimizing the reconstruction error between outputs $\{\hat E_i, \hat P_{xi}, \hat P_{yi}, \hat P_{zi}\}$ and inputs $\{E_i, P_{xi}, P_{yi}, P_{zi}\}$.}
    \label{fig:scheme_ae}
\end{figure}

Traditional density-based or distance-based anomaly detection methods don’t scale well with the size of the training set or the dimensionality of the input features. In contrast, deep neural networks succeed in modeling complex high-dimensional density distributions, and are thus used to process high-dimensional data. In particular, deep generative models have made great breakthroughs in generating complex distributions in the domain of computer vision and natural language modeling.
Deep generative models are trained to generate samples similar to the training samples, and are thus supposed to learn the data distribution and be able to evaluate the likelihood correctly.
It naturally leads to the solution of using deep generative models for new physics search, while taking in all low-level features as input and being as model-independent as possible. 
%Among the neural network based anomaly detection methods, reconstruction based anomaly detection, 
Especially, Autoencoders (AEs) and Variational Autoencoders (VAEs) have been explored for new physics searches recently.
%\cite{Heimel:2018mkt,Farina:2018fyg, Roy:2019jae, Blance:2019ibf}. Variational Autoencoders are also explored for event selection at the LHC \cite{Cerri:2018anq}.
While \cite{Cerri:2018anq, Blance:2019ibf} employ high-level features and physics observables as input, \cite{Heimel:2018mkt,Farina:2018fyg, Roy:2019jae} work on low-level jet constituents instead to tag non-QCD jets. For anomalous jet tagging, autoencoders trained with only QCD jets are used to detect non-QCD signals such as boosted top jets. In Fig. \ref{fig:scheme_ae}, we depict the schematic of AE-based anomalous jet tagger. Autoencoders employ a bottleneck architecture to effectively reduce the data dimensionality and extract relevant information for reconstructing input features.
%Using either images or four-vector-based input features, autoencoders have been shown to be potential in detecting heavy resonances \cite{Heimel:2018mkt,Farina:2018fyg}.
The basic idea of using autoencoders as anomaly detector is assuming that the trained AEs will be able to reconstruct samples similar to training samples (\textit{in-distribution}, InD), while giving large reconstruction errors when applied to unseen datasets (\textit{out-of-distribution}, OoD).  At the same time, adversarial training to decorrelate jet mass from the reconstruction error has been explored to assist bump-hunt based new physics searches \cite{Heimel:2018mkt}.
%Jet reconstruction is encouraged by minimizing the reconstruction error between input and output samples. 
%The reconstruction error is used as the anomaly score in most of the literature to date.

\begin{figure}[htb!]
    \centering
    \includegraphics[width=0.8\textwidth]{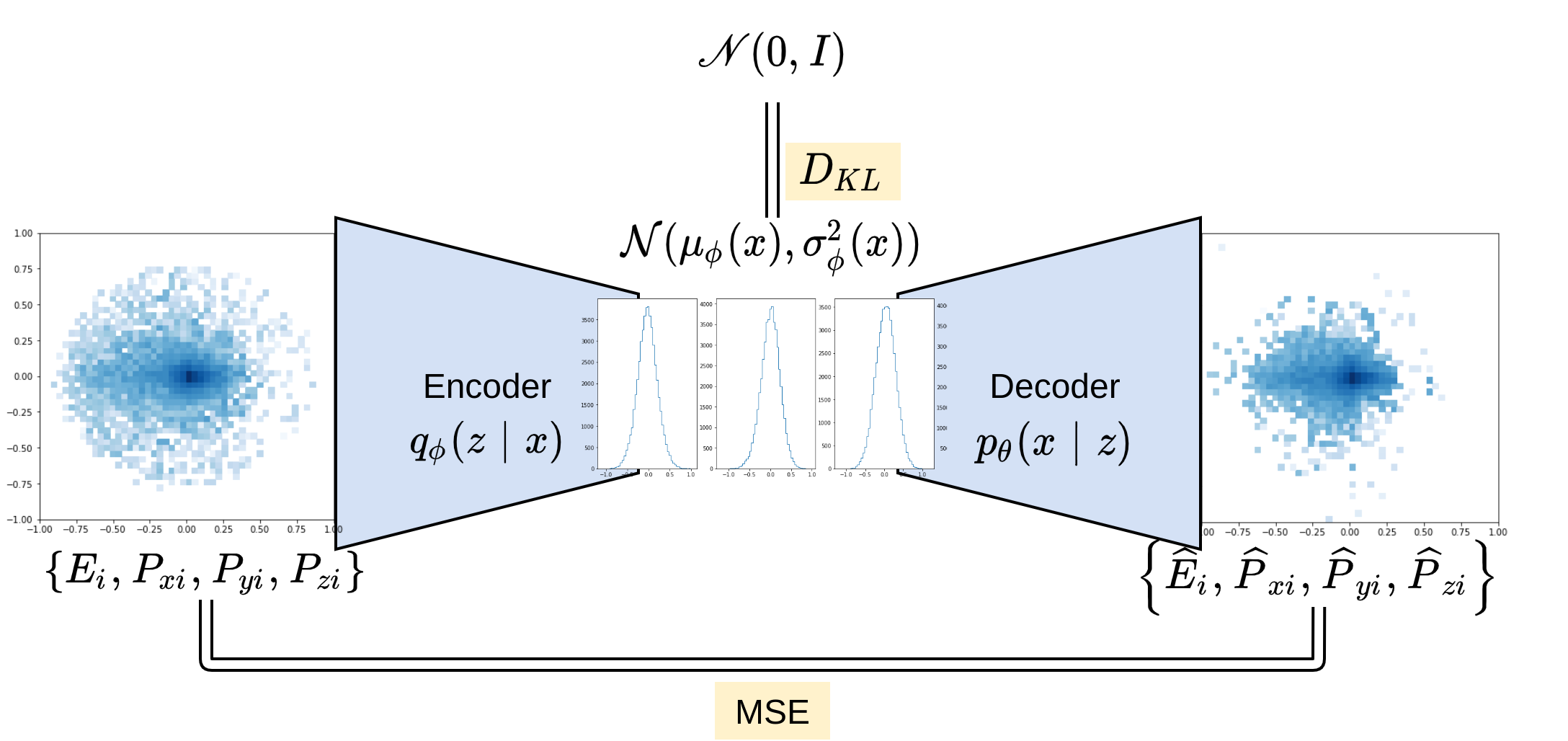}
    \caption{Schematic of Variational Autoencoder based anomalous jet tagger. Both the reconstruction error between outputs and inputs and the divergence of latent distributions are required to be optimized.}
    \label{fig:scheme_vae}
\end{figure}

% downsides and motivations
However, using reconstruction error to detect anomalies in this case is only an empirical, though sometimes effective, assumption. The latent representations of deterministic AEs are not optimized towards robust anomaly detection, and have random/unregularized behaviour (see the illustrative latent distribution of QCD samples in Fig. \ref{fig:scheme_ae}). 
%Regularization is used to restrict model complexity and thus improve generalization ability. The generalization ability is extremely important in model-independent signal detection.
A simple idea of regularizing the latent space will lead us to the Variational Autoencoders \cite{2013arXiv1312.6114K}. By modeling the distributions rather than the discrete points in the latent space, VAE promotes continuous latent representations capturing more physical meaning.
 As indicated in Fig. \ref{fig:scheme_vae}, VAE has a regularized latent distribution imposed by minimizing a divergence in the latent space, in addition to minimizing the reconstruction error in the input space as in the case of deterministic AEs.
Migrating from deterministic Autoencoders to Variational Autoencoders for anomalous jet tagging, there are a few motives: regularized latent representations, combination with generative modeling with latent variables, and a Bayesian framework for maximum likelihood estimation.

% core summary of what we have done.
While \cite{Cerri:2018anq} studies VAEs for event selection based on high-level observables, there is, however, not yet any application using VAEs in the context of low-level jet features.
In this work, we apply VAEs to anti-QCD tagging taking low-level constituents as input to maximize the information capacity, with possibilities of utilizing anomaly metrics in both input space and latent space.
%Input-space metric focuses on pixel-like space and have problems such as probability mis-assignment. Latent space L2 loss (is this connected with geometry of latent space).
%Based on the previously mentioned motivations and existing problems, we examine every building block in the process of using VAEs for anomalous jet tagging. 
%We first construct a basic VAE model for low-level constituents-based jet taggers.
%In order to ensure that we have a fully performing generative model, we examine its reconstruction performance, generation performance, and also properties of the latent space. 
To have a fair assessment of the anomaly detection performance, we tailor a series of test jet sets spanning in the range of different jet masses and jet types. Other than the reconstruction error used in corresponding deterministic AEs applications, we examine a few alternative anomaly metrics. To facilitate large radius jet based new physics searches such as bump-hunt, we implement a mass-decorrelated VAE tagger in a regularization-based approach \cite{Kasieczka:2020yyl}, which is faster and easier to train than an auxiliary adversarial network.

%There is another problem faced by generative model based anomaly detection.
Despite the fact that generative models provide us a simple approach for dealing with high-dimensional anomaly detection, they are not guaranteed to succeed for all the use cases. This problem is also reported in the computer vision applications of anomaly detection in the machine learning community.
In anomaly detection of natural images, it was found that sometimes higher probability is assigned to out-of-distribution samples than in-distribution samples \cite{2018arXiv181009136N, 2018arXiv181204606H}. 
%The pixel-space similarity is still not mature enough to render robust representation learning. And sometimes background confounding also brings spurious correlation that handicaps the OoD detection.
We observed the similar phenomenon in our anti-QCD tagging using VAEs.
%For instance, in the context of anti-QCD tagging, QCD jets might be assigned larger reconstruction error or lower likelihood than $W$ jets with two-prong structure, as will be shown later. 
%In the case some features dominate in the correlation with reconstruction error, anomalies in other domain would be able to show themselves effectively. %Specific mass-decorrelation methods may be applied to eliminate mass effects, such as adversarial training employed in \cite{Heimel:2018mkt}. The learnt information was not examined before and after adversarial training to answer the question of ``does limiting jet mass effects induce learning other information''.
%And other problem is effective OOD generalization: current methods generally tune hyper-parameters against a specific OOD dataset. This brings the under-optimality for other OOD types. To reduce this class-orientation and bring generalization across different OOD samples,
% oe-vae
Outlier Exposure (OE) \cite{2018arXiv181204606H} is proposed to solve this probability mis-assignment problem (i.e. outliers are assigned with higher probability). By injecting some outlier samples in the training process, it helps the autoencoder get better separation for InD samples and OoD samples. It is also reported that it generalizes to other OoD distributions. Furthermore, the auxiliary task of OE also provides us a very good handle to shape the latent data manifold tailored to our tasks. Thus we introduce the Outlier Exposed VAEs (OE-VAEs) for anomalous jet tagging in this work. In our case, we employ outlier exposure not only as an inducer for detecting sensitivity of unseen outlier samples, but also as a tool to guide and tune the information encoded. Using outlier samples as a leverage, we pursue decorrelation effects by using equivalent ``planing'' \cite{ATL-PHYS-PUB-2018-014, Bradshaw:2019ipy} (or, reweighting samples to obtain identical distributions) in jet mass. We achieve very good mass decorrelation by matching mass distributions between the exposed outlier dataset and the in-distribution training dataset, while at the same time gaining promising sensitivity for most jet types.

% organisation of the paper
The paper is organised as follows: in Section \ref{sec:vae} we present the basic settings of the problem, introduce the neural network architecture, and examine the properties of trained VAE models. 
In Section \ref{sec:anomaly_detection}, the performance in detecting non-QCD jets is carefully investigated. A comprehensive series of test sets are generated to fully examine the anomaly tagging performance in different jet types. At the same time, the baseline mass-decorrelated taggers with distance correlation as the regularizer are introduced and studied. OE-VAEs, the outlier exposure solution to improve sensitivity and achieve mass-decorrelation, is presented in Section \ref{sec:oe}. Finally we summarize this work in Section \ref{sec:summary}.

%%%%%%%%%%%%%%%%%%%%%%%%%%%%%%%%%%%%%%%%%%%%%%%%%%%%%%%%%%%%%%%%%
\section{\label{sec:vae}Variational Autoencoders for Anomalous Jet Tagging}

We provide here some mathematical foundation of VAEs \cite{2013arXiv1312.6114K, 2016arXiv160605908D}.
% variational inference
%Practically speaking, a VAE, similar to an AE, consists of an encoder and a decoder, which are both parameterized with neural networks. 
By parameterizing the variational inference and the generative model with deep neural networks, VAEs map input data $x \in \mathcal{X}$ into the latent space $z \in \mathcal{Z}$ with $q(z|x)$ and map the latent representation $z$ back into the input space with $p(x|z)$. As briefly mentioned in the introduction, VAEs can be viewed as the regularized version of deterministic AEs by imposing latent structure with a Kullback--Leibler divergence (KL divergence, $D_{\rm KL} [p(x) \Vert q(x)] = \int dx~p(x) \log \frac{p(x)}{q(x)}$) from the prior distribution to the posterior distribution for latent variables. The training objective of VAEs includes an extra term of minimizing the KL divergence between the prior $p(z)$ and posterior $q(z|x)$ distributions of latent variables $z$.
%Despite the fact that VAE is a deep generative model, it trains fast by simple back-propagation with a reparameterization trick. The reparameterization instead reparameterizes the latent sampling using an extra input layer by the change of variable $z=\mu + \sigma \odot \epsilon$ (with $\epsilon \sim \mathcal{N}(0, I)$), in order to back-propagate through the stochasticity of latent sampling. 

The actual objective of the VAE is to approximately maximize the log-likelihood $\log p(x)$, within a Bayesian inference framework.
%variational distribution $q(z|x)$ for latent variables
The log-likelihood of the input data distribution can be written in terms of:
\begin{equation}
\label{eqn:elbo}
    \log p(x) - D_{\rm KL}[q(z \vert x) \Vert p(z \vert x)] = \mathbb{E}_{z\sim q(z \vert x)} [\log p(x \vert z)] - D_{\rm KL}[q(z \vert x) \Vert p(z)] \, ,
\end{equation}
in which $\mathbb{E}_{z\sim q (z \vert x)} [\log p(x \vert z)]$ is empirically the negative reconstruction error of an autoencoder. The left-hand side of Eq. \ref{eqn:elbo} is the marginal log-likelihood we want to maximize (the other divergence term can be effectively minimized in the case of a powerful encoder $q(z \vert x)$). The right-hand side of Eq. \ref{eqn:elbo} is called the \emph{Evidence Lower BOund} (ELBO), since it gives a lower bound of the log-likelihood. Negative ELBO serves as the minimization objective in the VAE training , and is practically the sum of the reconstruction error of the autoencoder and the divergence between the posterior and prior distributions in the latent space. Thus the empirical training objective of VAE can be written as:
\begin{equation}
\label{eqn:loss_vae}
   \mathcal{L}_{\rm VAE} = \mathcal{L}_{\rm recon} + \mathcal{L}_{\rm KL} \, .
\end{equation}
%From another point of view, VAEs can be seen as a regularized version of deterministic autoencoders. By imposing explicit constraints in the form of the latent prior distribution, we have a handle of how the latent variables behave. Then the training of VAEs includes matching the posterior to the prior distribution by minimizing the KL-divergence from prior to posterior distributions in latent space. 

% inference and generative
%The encoder and decoder within a VAE serve as the inference network and the generative network respectively. The latent space from the encoding process provides the posterior latent representation. 
In standard VAEs, prior latent distributions are assumed to be multivariate standard Gaussian distribution $\mathcal{N}(0, I)$. And posteriors are estimated in the form of $\mathcal{N} (\mu(x), \sigma^2 (x))$ by mapping input data points to means $\mu(x)$ and variances $\sigma^2(x)$ of  Gaussian distributions. The latent variables $z$'s are then sampled from this posterior and mapped back into the input space $\mathcal{X}$ via the decoder $p(x \vert z)$.
%\footnote{Mixed latent models might help with clusterability.}
On the other hand, sampling from the latent distribution $\mathcal{N}(0, I)$ will facilitate generation of new samples as long as we have trained the VAE well and obtained a powerful decoder. So examining the quality of the generated samples also serves as an important measure on how well the VAE is learning the input data distribution.

As introduced previously, KL divergence in the VAE objective can be viewed as a regularization term. VAEs with variable regularization weight other than 1 are also formulated as $\beta$-VAE \cite{Higgins2017betaVAELB} models.
%The problem of latent space in Autoencoders: \mynote{plot for comparison in latents}
To clarify this process, the empirical objective of a $\beta$-VAE is written as:
\begin{equation}
    \mathcal{L}_{\rm \beta-VAE}= - \mathbb{E}_{z\sim q(z \vert x)} [\log p(x \vert z)] +
    \beta D_{\rm KL} [q(z \vert x) \Vert p(z)]  = \mathcal{L}_{\rm recon} + \beta \mathcal{L}_{\rm KL}~ ,
\end{equation}
where $\beta$ denotes the relative strength of the latent regularization. Changing $\beta$ affects the competition between fitting the latent distribution and the input space reconstruction.  With $\beta=0$ the VAE reduces to the deterministic autoencoder, for which only the reconstruction is optimized during training, resulting in very good jet reconstruction, however, losing inference capability in the latent space. Increasing $\beta$ leads to compensation between the jet reconstruction and the latent distribution matching.
% extensive arguments on beta-VAE
%On the other hand, $\beta-$VAE was shown to have disentanglement effects in latent space. 
We have tested different values $\beta=0.1, ~0.5, ~1.0, ~5.0$, and observed that  $\beta=0.1$ gives better balance between input reconstruction and latent coding.
We focus on reporting results for $\beta=0.1$ in this work. In the following text, we use the notations of VAE and $\beta$-VAE interchangeably unless stated specifically.

\subsection{Neural Network Architecture}

\paragraph{Datasets}

We train on simulated QCD jets collected from the fatjet trigger criteria of ATLAS Collaboration.
\footnote{https://twiki.cern.ch/twiki/bin/view/AtlasPublic/JetTriggerPublicResults} 
QCD di-jet events are generated with MadGraph \cite{Alwall_2011} for LHC 13 TeV, followed by Pythia8 \cite{Sj_strand_2008} and Delphes \cite{de_Favereau_2014} for parton shower and fast detector simulation, respectively. No pile-up was simulated.
All jets are clustered using the anti-$k_T$ algorithm \cite{Cacciari_2008} with cone size $R=1.0$ and the selection cut $\pt > 450$ GeV. Particle flow objects are used for jet clustering, with no jet trimming applied.

\paragraph{Input features and Preprocessing} We take the first 20 \footnote{The number of input jet constituents 20 is chosen to optimize the signal significance of boosted top jets.} hardest ($p_{\rm T}$ ordered) jet constituents as inputs in the format of four vectors $\{x_i = (E_i, P_{xi}, P_{yi}, P_{zi}); i = 1, ..., n\}$ with $n=20$. 
Jets are preprocessed with minimum transformation to avoid designing bias. Jets are longitudinally boosted and rotated to center at $(0,0)$ in the %\textcolor{red}{$(y, \phi)$}
$(\eta, \phi)$ plane. Centered jets are then rotated so that the jet principal axis $(\sum_i \frac{\eta_i E_i}{R_i}, \sum_i \frac{\phi_i E_i}{R_i})$ (with $R_i = \sqrt{\eta_i^2 + \phi_i^2)}$) is vertically aligned on the $(\eta, \phi)$ plane, with the rotation angle $\alpha$ indicated in Eq. \ref{eqn:principal_axis}: 

\begin{equation}
\label{eqn:principal_axis}
    \tan \alpha = \frac{\sum_i \frac{\phi_i E_i}{R_i}}{\sum_i \frac{\eta_i E_i}{R_i}} \, .
\end{equation}

We standardize
%\footnote{Using the \emph{RobustScaler} in \emph{scikit-learn} \cite{scikit-learn}} 
the input features before feeding into the VAEs, by removing the median and scaling with the interquartile range. 

\paragraph{VAE Architecture}
% VAE architecture and parameters
We explored two simple architectures: Fully Connected Networks (FCN) and Long Short-Term Memory (LSTM) Networks. Since no signiﬁcant difference in performance was observed, we only present the results for FCN based VAEs (FCN-VAEs).

\begin{figure}[htb!]
    \centering
    \includegraphics[width=0.8\textwidth]{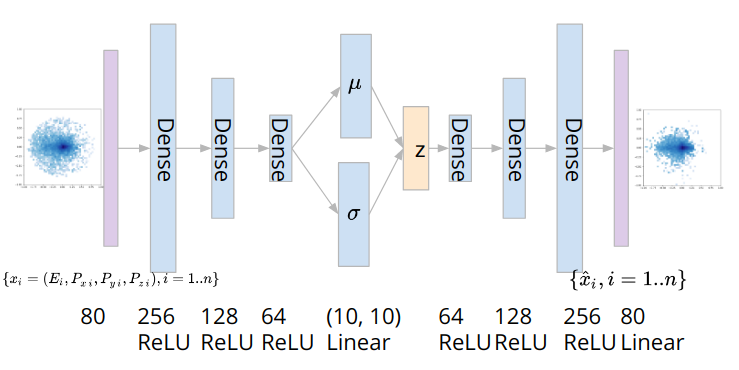}
    \caption{Architecture of FCN-VAE, with an input dimension of 80 and a latent dimension of 10. Dense layers with ReLU activations are employed for the encoder and decoder. Linear layers are used to model the latent Gaussian means and logarithms of variances. Output features are generated through a linear layer with the same dimension as the input layer.}
    \label{fig:vae_fcn}
\end{figure}

For FCN-VAEs, simple dense layers are employed for the encoder and the decoder. ReLU activations are used through latent layers, and linear activation is used in the output layer. The encoder and decoder have symmetric architectures, as the encoder is composed of 256, 128, 64 neurons for each hidden layer. The latent dimension of 10 has been optimized to maximise boosted top signal significance. The latent Gaussian means $\mu$ and logarithms of variances $\log(\sigma^2)$ are parametrized by linear dense layers. 
Then the latent vector $z$ sampled from the posterior distribution $\mathcal{N}(\mu, \sigma^2)$ is passed to the decoder to reconstruct the output jet. Generated features are read out through a linear output layer with the same dimension as the input layer.
A brief summary of the VAE architecture is depicted in Fig. \ref{fig:vae_fcn}.
The VAE loss is written as in Eq. \ref{eqn:vae_loss}.
The reconstruction error is simply chosen as the Mean Squared Error (MSE) between input features $\{x_i; i = 1...n\}$ and output features $\{\hat x_i; i = 1...n\}$.

\begin{equation}
\label{eqn:vae_loss}
    \mathcal{L}_{\rm \beta-VAE} =  \mathcal{L}_{\rm recon} + \beta \mathcal{L}_{\rm KL} =  \frac{1}{n} \sum_{i=1}^n \lVert \hat x_i  - x_i \rVert^2 + \beta D_{\rm KL}[q(z|x) \Vert p(z)]
\end{equation}

%\subsection{Regularization in Latent Space}
%\paragraph{Training Tricks}
%\begin{itemize}
%    \item annealing or cyclic annealing \cite{abs-1903-10145}: slowly increasing regularization strength in the training process.
%    \item mixing samples (exposing anomalous class samples/ classification as auxiliary task)
%\end{itemize}

\paragraph{Training Setup}

Throughout this study, we train on 600,000 QCD jets, of which 20\% serves as the validation set. 
%\footnote{\mynote{The QCD dataset can be found here. (zenodo)}} 
We employ the Adam \cite{kingma2014adam} algorithm for optimization, with the default parameters and a learning rate of 1e-3. VAEs are trained for 50 epochs with a batch size of 100. 
%We haven't done a comprehensive hyperparameter optimization. This is left for future work.
%We applied annealing training \cite{} to avoid non-informative latent codes.

\subsection{Examining trained VAEs}
% what properties to be examined and the motivation: 
We examine a few properties of the trained VAE models: jet reconstruction, jet generation, and latent representations.

\paragraph{Jet Reconstruction}

\begin{figure}[htb!]
    \centering
    \includegraphics[width=0.45\textwidth]{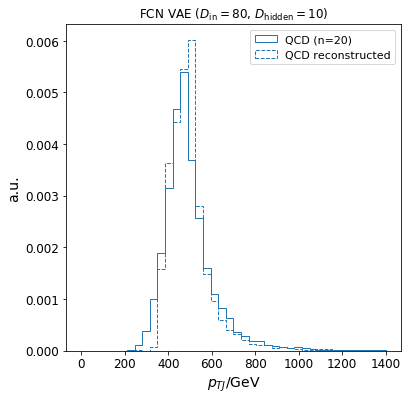}
    \includegraphics[width=0.45\textwidth]{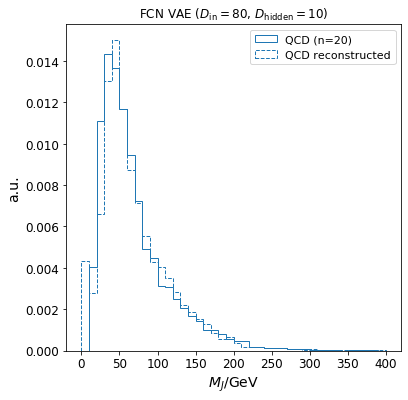}
    \caption{Reconstructed (\textit{dashed lines}) jet $p_{\rm T}$ and $M_{\rm J}$, compared with input jet distributions (\textit{solid lines}) plotted for the first 20 jet constituents with $\pt$-ordering.}
    \label{fig:vae_recon}
\end{figure}

To examine how well QCD jets can be reconstructed by VAEs, we show a few distributions of reconstructed high-level jet features in Fig. \ref{fig:vae_recon}. We see that jet $p_{\rm T}$ and mass are both well reconstructed.

\paragraph{Jet Generation}

\begin{figure}[htb!]
    \centering
    \includegraphics[width=0.45\textwidth]{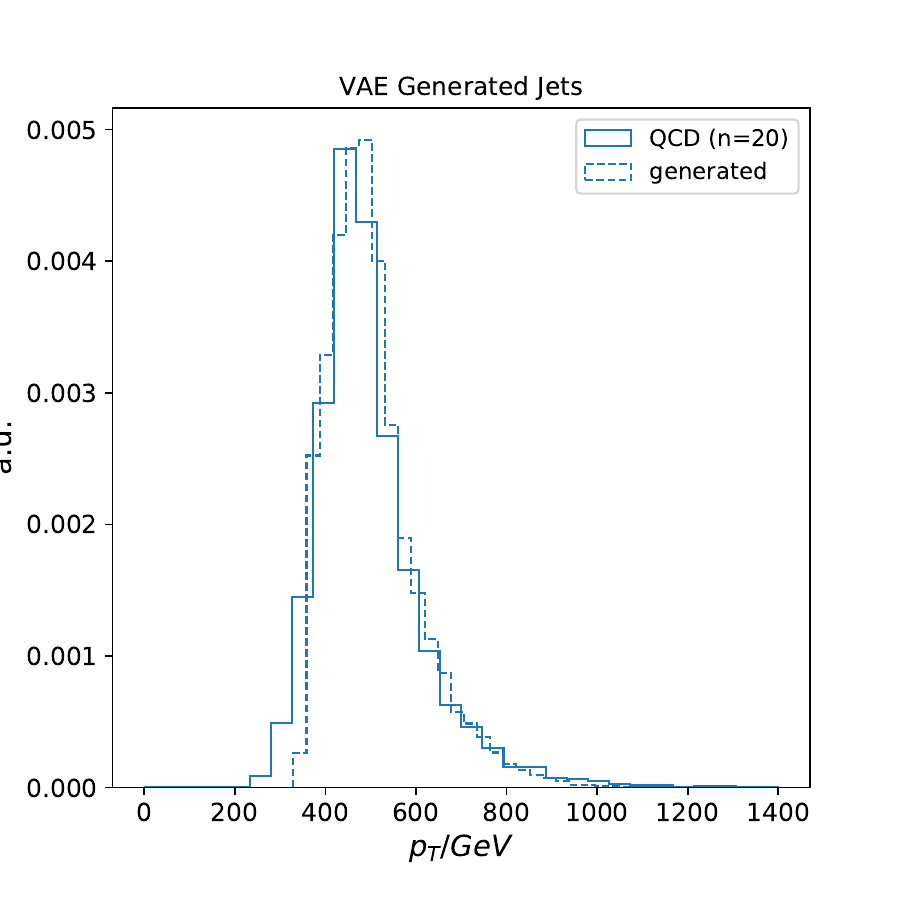}
    \includegraphics[width=0.45\textwidth]{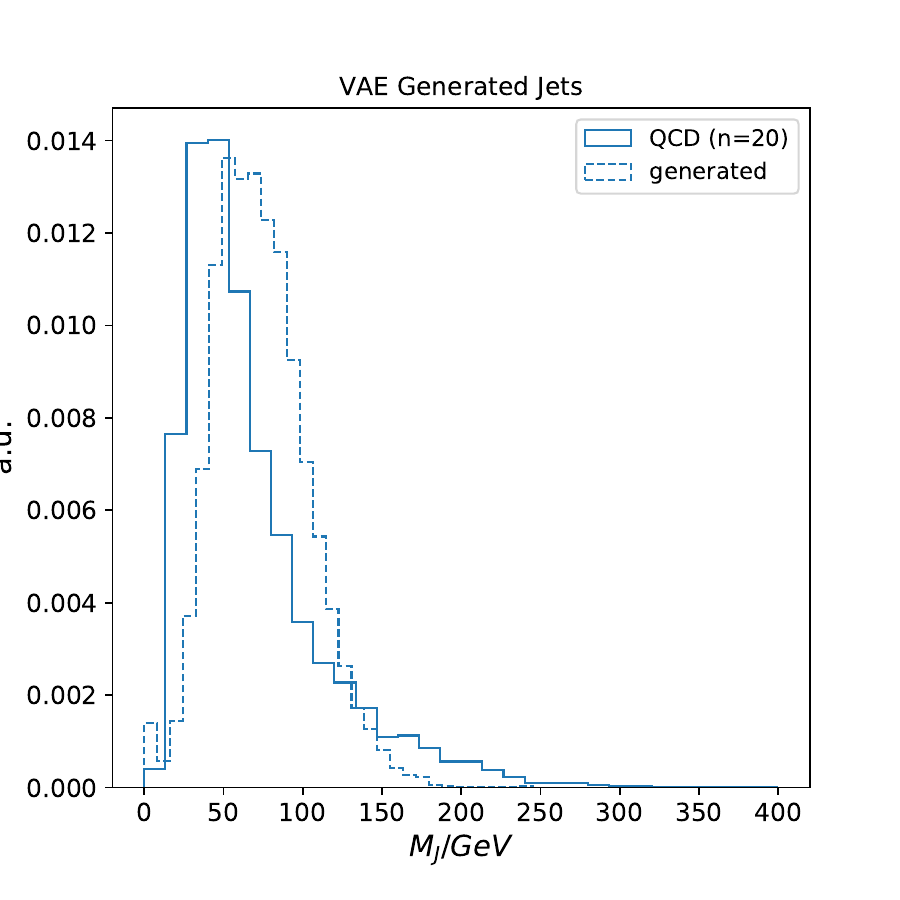}
    \caption{High-level feature distributions of VAE generated jets (\textit{dashed lines}) by sampling from the latent distribution $z \sim \mathcal{N} (0, I)$. \textbf{Left}: $p_{\rm T}$; \textbf{Right}: $M_{\rm J}$.}
    \label{fig:generation_obs}
\end{figure}

To assess the quality of the generative modeling, we sample from the prior distribution $z \sim \mathcal{N} (0, I)$ in the latent space and generate output jets by passing the samplings through the decoder. High-level observables constructed for the generated jets $p_{\rm T}$ and $M_{\rm J}$ are shown in Fig. \ref{fig:generation_obs}. Both generated $p_{\rm T}$ and $M_{\rm J}$ distributions match well with the ones of the original QCD dataset. However, there is a small discrepancy for jet mass, suggesting that the model architecture could be further optimized.

\paragraph{$\beta$-VAE Latent Representation}

\begin{figure}[htb!]
    \centering
    \includegraphics[width=0.45\textwidth]{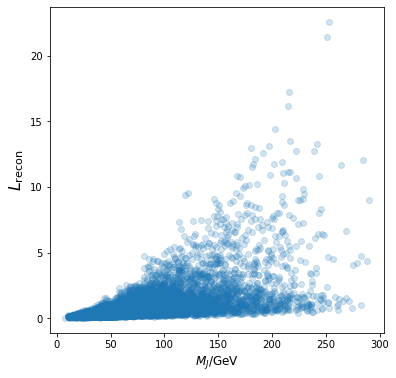}
    \includegraphics[width=0.45\textwidth]{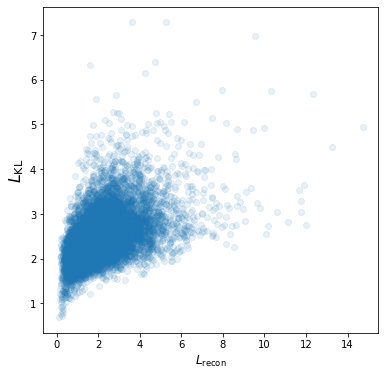}
    \includegraphics[width=0.45\textwidth]{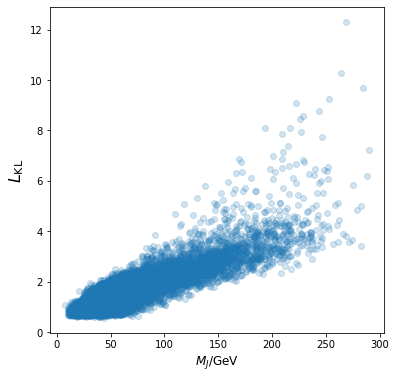}
    \includegraphics[width=0.45\textwidth]{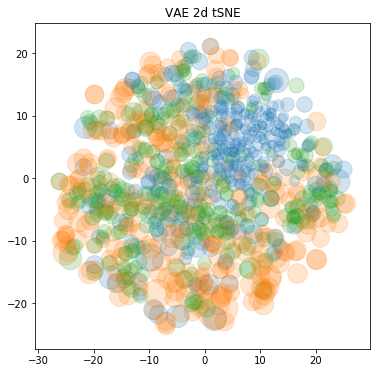}
    \caption{\textbf{Upper-Left}: Reconstruction error v.s. jet mass; 
    \textbf{Upper-Right}: KL divergence v.s. reconstruction error;
    \textbf{Lower-Left}: KL divergence v.s. jet mass;
    \textbf{Lower-Right}: Latent 2d-tSNE visualisation for different jet types (QCD jets (\texttt{Blue}), W jets (\texttt{Green}) and top jets (\texttt{Orange})). All plots are for FCN-VAE with $\beta=0.1$.}
    \label{fig:beta0.1_latent_examine}
\end{figure}

As discussed previously, $\beta$ controls the balance between input reconstruction and latent coding. When $\beta$ increases, the reconstruction performance might drop since there is extra effort to fit the latent distribution. When $\beta$ is small enough, it approaches the behaviour of deterministic autoencoders. In Fig. \ref{fig:beta0.1_latent_examine}, model examination results for $\beta=0.1$ are shown.
We plot the correlation between the reconstruction error and the jet mass, the correlation between the latent KL divergence and the jet mass, the correlation between the reconstruction error and the latent KL divergence, and the 2d-tSNE \cite{vanDerMaaten2008} (\texttt{perplexity} $=50$) visualisation of the latent representations for in-distribution QCD jets (\texttt{Blue}), out-of-distribution $W$ jets (\texttt{Green}), and out-of-distribution top jets (\texttt{Orange}). 
%Size of dots are corresponding KL divergence.
We observe that the reconstruction error has an upper-bounded correlation with the jet mass. And the latent KL divergence is also (even more strongly and linearly) correlated with the jet mass. This suggests that the regularized latent space has encoded relevant information but has different geometry w.r.t. the input space.
Little clustering effect is present in the 2d-tSNE visualisation. As $\beta$ increases (see App. \ref{app:betavae}), a stronger clustering effect is observed at the cost that reconstruction is less maintained.  

%\begin{figure}
%    \centering
%    \includegraphics[width=1.0\textwidth]{figs/vae_latents_sample.png}
%    \caption{Caption}
%    \label{fig:beta0.1_latents}
%\end{figure}

%\paragraph{Latent Representation -- kNN outlier detection}

\section{\label{sec:anomaly_detection}Performance in Anomalous Jet Tagging}

In this section, we present the performance of VAE-based anomalous jet taggers, tested on different jet types.
%, and \alarm{the responses on jet kinematics}. 
Receiver Operating Characteristic (ROC) curves and the area under the ROC Curve (AUC) are used for measuring the performance universally.

\paragraph{Test Datasets}

A series of test sets \cite{cheng_taoli_2020_3901833} are generated to fully examine the performance of VAE-based anti-QCD jet tagging. 
Boosted $W$ jets, top jets, and Higgs jets are generated as representatives of two-prong, three-prong, and four-prong jets.
To test jet mass effects, we composed $W$ jet datasets with rescaled $W$ jet masses (Standard Model $W$ jets with only the mass changed for event generation) for comparison. The same mass-rescaling strategy is also applied to top jets.
We employ Two Higgs Doublet Models (THDMs) \cite{Branco:2011iw} for generating boosted Higgs jets.
Heavy Higgs bosons are generated in pairs ($p p \to H H$), with the requirement of $p_{\rm T} > 600~\rm{GeV}$ and decaying into light Higgs pairs $H \to h h$. The light Higgs bosons are then restricted to the $h \to b \bar b$ decay mode.
Different light Higgs masses are experimented to show different degrees of ``four-prongness". With a very light $m_h$, boosted heavy Higgs jets will resemble two-pronged jets.
All sample jets are clustered using the anti-$k_T$ algorithm with a cone size of $R=1.0$. When testing on these datasets, jet $p_{\rm T}$s are restricted to [550, 650] GeV for a fair comparison. The test set size for each jet category is set to be 20,000.

Here is the detailed information for the test sets:
\begin{itemize}
    \item \textit{Two-prong\hspace{0.5cm}} boosted $W$ jets are produced by the decay of a heavy resonance $W'$ with $m_{W'} = 1.3$ TeV, with $W$ decaying to light quark jets and $Z$ decaying to neutrinos: $ pp \to W' \to W(jj) Z(\nu \nu)$. Masses experimented include $m_W =  59, ~ 80, ~120, ~174$ GeV.
    \item \textit{Three-prong\hspace{0.5cm}} boosted top jets are generated with the decay of a heavy resonance $Z'$ of $m_{Z'}=1.5$ TeV:  $ pp \to Z' \to t \bar t$. ``Top'' masses are set to be $m_t = 80, ~174$ GeV. (for $m_t = 80$ GeV, the decay product $W$ mass is set to be 20 GeV).
    \item \textit{Four-prong\hspace{0.5cm}} boosted heavy Higgs pair production in THDMs is borrowed to generate four-prong samples: $pp \to HH$, with $H \to h(bb) h(bb)$ and $m_H = 174$ GeV, $m_h = 20, ~80$ GeV. In the event generation, we employ $h_3$ 
    (with $p_T > 600$ GeV)
    in THDM as the heavy Higgs, and $h_1$ as the light Higgs. 
    %\alarm{link with h3h1 in plots legend}
\end{itemize}

\paragraph{Anomaly Metric}

After successfully training the VAEs, utilizing the trained models as effective anomalous jet taggers also requires a good anomaly metric well-defined and specifically tailored.
% motivations for these options
%Even though being the most widely used anomaly metric in the literature, the naive MSE is confronted with a few problems: there is some correlation with jet constituent numbers, although in fixed input length this effect would be eliminated somehow; and strong mass correlation is generally present. 
%Such effects might be better controlled using an anomaly metric based on the low-dimensional latent representations since the data geometry is expected to be much simpler in that space.
MSE is the most widely used anomaly score in the literature. For VAEs, there are more options since the latent space is now serving as a regularized abstract space for relevant physics information.
We thus explored how the KL divergence from the prior to the posterior distribution in the latent space works as an anomaly metric. Other than that, the negative log-likelihood (NLL) is also directly considered as an anomaly score. 
Exploring other ``similarity'' measures in the input space, an optimal transport based metric \cite{Komiske:2019fks} measuring similarities between input jets and reconstructed jets is also tested as an anomaly score. 
%It explores in the input space which similarity measure brings better discrimination ability.
Besides, independently from the VAE setting, we found that a simply defined statistical anomaly score for low-level input features already works well, given proper preprocessing and standardization of the input features.

%\alarm{With proper preprocessing and standardization, it effectively builds a Gaussian model for input features.}

%Other than that, a simple kNN-based outlier detector was also investigated to manifest the learned latent representations.
%\mynote{[plot: MSE investigation plots]; maybe put busy ROC curves in appendix}
Here is a summary of the options investigated as anomaly metrics:

\begin{itemize}
    \item Negative log-likelihood: $\mathcal{L}_{\rm VAE} = \mathcal{L}_{\rm recon} + \mathcal{L}_{\rm KL}$. 
    %\mynote{is the loss exactly the nll? No, beta-VAE loss is not exactly the NLL.}
    \item MSE reconstruction error in the input space: $\mathcal{L}_{\rm recon} = \frac{1}{n} \sum_i \Vert \hat x_i - x_i \Vert^2$. 
    \item KL divergence in the latent space: $\mathcal{L}_{\rm KL} = D_{\rm KL}[q(z|x) \Vert p(z)]$.
    \item Energy Mover's Distance (EMD): EMD is defined as a metric in the collider space using optimal transport to find the minimum energy moving strategy between two LHC events. The EMD between event $\mathcal{E}$ and $\mathcal{E'}$ is defined as:
    \begin{equation}
    \label{eqn:emd}
        {\rm EMD} (\mathcal{E}, \mathcal{E'}) =      
        \min_{f_{ij}}
        \sum_{ij} f_{ij} \frac{\theta_{ij}}{R} + \left \vert \sum_i E_i - \sum_j E_j' \right \vert, 
    \end{equation}
    \begin{equation*}
        f_{ij} \geq 0, \sum_j f_{ij} \leq E_i, \sum_i f_{ij} \leq E_j', \sum_{ij} f_{ij} = E_{\min},
    \end{equation*}
    where $\theta_{ij}$ is the angular distance between particles indexed with i and j in $\mathcal{E}$ and $\mathcal{E'}$, and $f_{ij}$ denotes the energy being moved between events. $E_{\min} = \min \{\sum_i E_i, ~ \sum_j E_j'\}$. $R$ is a weight parameter and is set to be 1.0 in our practice. We calculate EMDs between input jets and output jets as the anomaly score. We expect out-of-distribution jets will give larger EMDs in the sense that they will not be easily reconstructed. We have already normalized jet $p_{\rm T}$s to eliminate the contribution from pure energy difference in the second term of Eq. \ref{eqn:emd}.  
    \item MSS: the simple mean squared score (MSS) after standardizing input features: $\frac{1}{n} \sum_i \Vert x_i \Vert ^2$.
    %\footnote{If using the original unscaled data this is $\frac{1}{n} \sum_i \Vert \frac{x_i-\tilde{x_i}}{\textrm{IQR}_i} \Vert ^2$ with $\tilde{x_i}$ denoting the median and IQR denoting the interquartile range.}
    This can be seen as an equivalent of the $\chi^2$ statistic of Gaussian distributed data. 
    %$\frac{1}{n} \sum_i \Vert \frac{x_i-\tilde{x_i}}{\textrm{IQR}_i} \Vert ^2$
    %\item \emph{Reconstruction Probability}
    %\item kNN Outlier 
\end{itemize}

\subsection{Results}

\begin{figure}[htb!]
    \centering
    \includegraphics[width=0.6\textwidth]{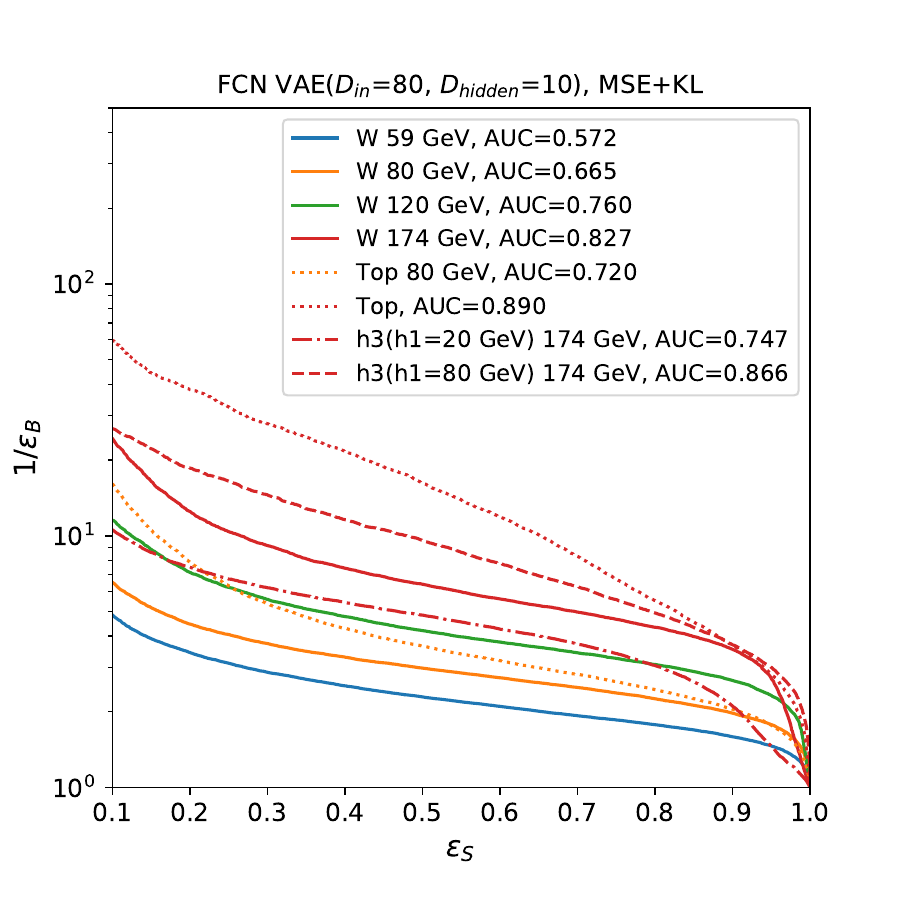}
    \caption{ROC curves for the spectrum of test signal samples, with NLL as the anomaly score. Colors are used to denote jet masses, while different line styles are used for different jet types. We have solid lines for $W$ jets of four different masses, and \textcolor{red}{\texttt{red}} indicates the benchmark mass of 174 GeV for any jet type.}
    \label{fig:vae_rocs_elbo}
\end{figure}

We first present the ROC curves for NLL-based anomaly score. This is aiming at seeing the general responses on different jet masses and types. In Fig. \ref{fig:vae_rocs_elbo}, ROC curves under NLL are presented for the full spectrum of test signals. 
%To see more clearly how the VAE performance is affected by jet masses, we plot ROC curves for $W$ jets with variable masses in the left panel of Fig. \ref{fig:vae_rocs_elbo_split}.
From the plot, it's obvious that the VAE performance in AUC is correlated with the jet mass. For mass-rescaled $W$ jets and top jets, they all show the same trend. As for the jet complexity, we take the jet mass of 174 GeV as a benchmark and show the ROC curves for 174 GeV jets in \textcolor{red}{\texttt{red}} lines.  %in the right panel of Fig. \ref{fig:vae_rocs_elbo_split}  
We observe that top jets have the highest discriminative scores, while $W$ and Higgs jets have slightly lower AUCs. \footnote{This might be because we tuned a few hyper-parameters such as number of input constituents according to the top test set.}
 In the rest frame of the heavy Higgs, light Higgs jets with a very small mass are almost produced back to back and are very boosted. Thus \texttt{h3} with the decaying product \texttt{h1} of a mass of 20 GeV should behave similarly to two-prong jets.
In Fig. \ref{fig:vae_rocs_elbo}, \texttt{h3(h1=20GeV)} has a lower AUC w.r.t. \texttt{h3(h1=80GeV)}, as expected in the sense that \texttt{h3(h1=20GeV)} has a simpler two-prong-like substructure.

\begin{figure}
    \centering
    \includegraphics[width=0.9\textwidth]{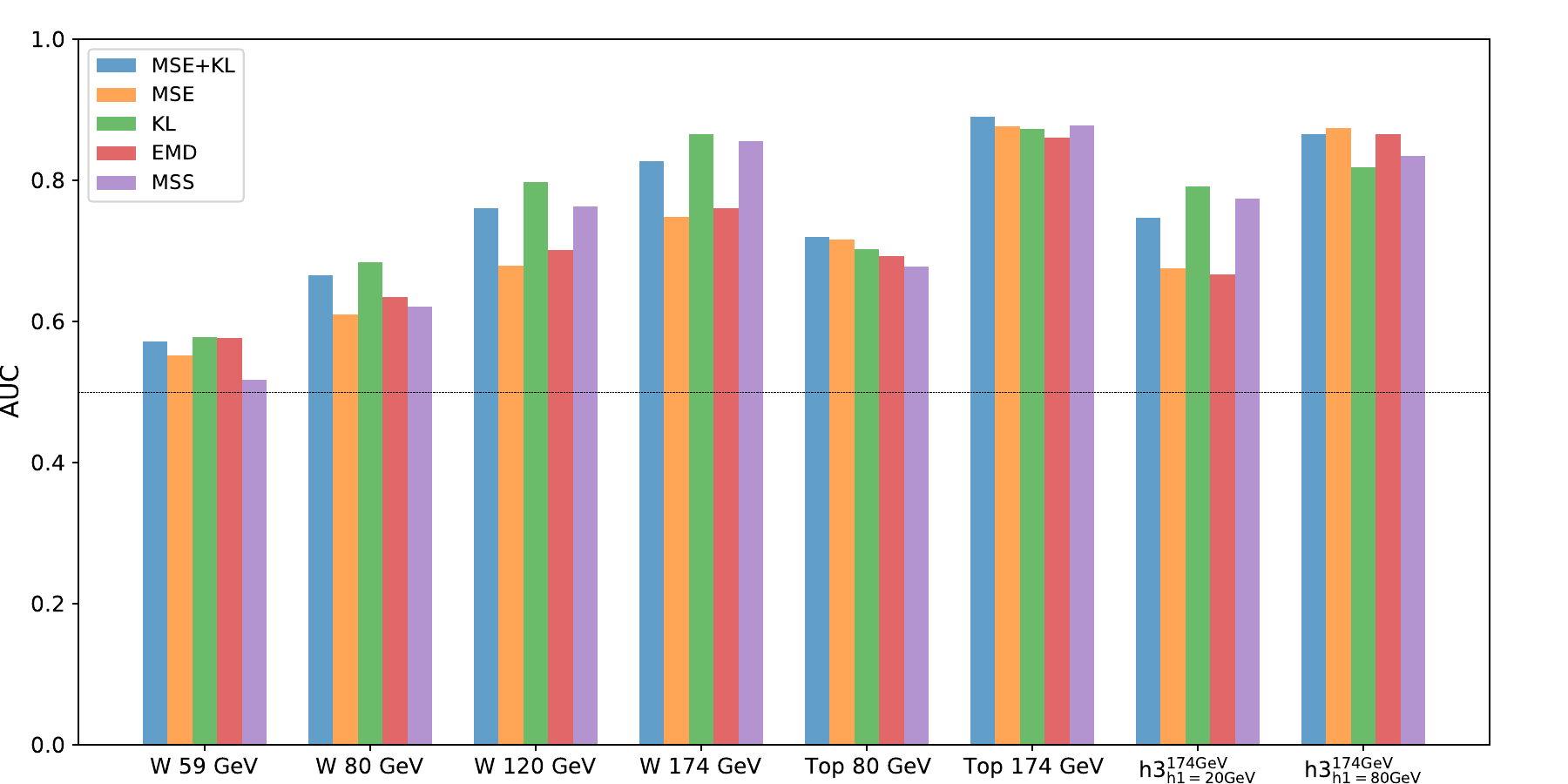}
    \caption{Summary plot of AUCs in the complete spectrum of test signal samples, for all the anomaly metrics investigated.}
    \label{fig:aucs_vaemodels}
\end{figure}

\begin{table}[htb!]
    \centering
    \begin{tabular}{c|cccccccc}
    \hline
    \multirow{2}{*}{\backslashbox{Metric}{Signal}}  &  \multicolumn{4}{c}{$W$}  & \multicolumn{2}{c}{Top} & \multicolumn{2}{c}{Higgs}\\ 
    & $W_{59 \textrm{GeV}}$ &  $W_{80 \textrm{GeV}}$ & $W_{120 \textrm{GeV}}$ & $W_{174 \textrm{GeV}}$ & Top$_{80 \textrm{GeV}}$ & Top$_{174 \textrm{GeV}}$ & h3$^{174 \textrm{GeV}}_{\textrm{h1=20GeV}}$ & h3$^{174 \textrm{GeV}}_{\textrm{h1=80GeV}}$\\
    \hline \hline
    MSE+KL  &    0.572  & 0.665  & 0.760  & 0.827  & \textbf{0.720}  & \textbf{0.890}  & 0.747  & 0.866 \\
    % wrong m80 top
    %MSE+KL  & 0.573 & 0.664 & 0.760 & 0.827 & 0.800 & \textbf{0.890} & 0.744 & 0.866 \\  
    % hidden data
    %MSE+KL(noannealing, epochs=50) &   0.5681309037500001  & 0.6534394112499999  & 0.7498504724999999  & 0.82294649  & 0.7983469106162937  & 0.88818884375  & 0.7427797225  & 0.8660658699999999\\
    %MSE+KL(annealing, epochs=50) &  0.572080565  & 0.6564554362500001  & 0.75437117375  & 0.8264171137499999  & 0.8028420707577038  & 0.8879099450000001  & 0.7457844150000001  & 0.8636024474999999\\
    %MSE+KL(epochs=100)  & 0.573  & 0.661  & 0.757  & 0.827  & 0.800  & 0.889  & 0.742  & 0.867  \\
    %MSE+KL(epochs=200) & 0.559 & 0.659  & 0.764  & 0.832 & 0.802  & 0.889  & 0.747  & 0.864 \\
    %MSE+KL(epochs=500) & 0.569  & 0.670  & 0.767  & 0.832  & 0.801  & 0.891 & 0.744  & 0.865 \\
    MSE     & 0.552  & 0.610  & 0.679  & 0.748  & 0.716  & 0.877  & 0.675  & \textbf{0.874} \\
    KL      & \textbf{0.578}  & \textbf{0.684}  & \textbf{0.797}  & \textbf{0.866}  & 0.702  & 0.873  & \textbf{0.791}  & 0.818 \\
    % wrong m80 top
    %MSE     & 0.553 & 0.609 & 0.679 & 0.748 & 0.746 & 0.877 & 0.672 & \textbf{0.874} \\
    % wrong m80 top
    %KL      & \textbf{0.578} & \textbf{0.684} & \textbf{0.797} & \textbf{0.866} & \textbf{0.829} & 0.873 & \textbf{0.789} & 0.819 \\
    EMD     & 0.576  & 0.635  & 0.701  & 0.761  & 0.692  & 0.860  & 0.667  & 0.865\\
    % wrong m80 top
    %EMD     & \textbf{0.578} & 0.634 & 0.701 & 0.761 & 0.734 & 0.860 & 0.664 & 0.865 \\
    MSS     & 0.517  & 0.621  & 0.763  & 0.855  & 0.678  & 0.878  & 0.774  & 0.835\\
    % wrong m80 top
    %MSS     & 0.517 & 0.621 & 0.763 & 0.855 & 0.822 & 0.878 & 0.772 & 0.836 \\
    \hline
    \end{tabular}
    \caption{AUCs for VAE tested on the complete spectrum of test signal samples. Results for different metrics are presented for comparison. The highest AUCs in each signal are highlighted.}
    \label{tab:aucs_vae}
\end{table}

% anomaly metrics
Then we focus on examining the performance of different anomaly scores. In Fig. \ref{fig:aucs_vaemodels}, we collectively show AUCs for all the anomaly metrics introduced previously. Meanwhile, a complete summary of AUCs is recorded in Table \ref{tab:aucs_vae}. (The full ROC curves can be found in App. \ref{app:rocs}.)

In general, the results from different anomaly scores support each other. The mass correlation trend holds in all the cases. 
The two input space reconstruction based scores \texttt{MSE} and \texttt{EMD} perform similarly. This makes sense since they both measure the similarity in the input space. However, their performance is worse than the other scores in most of the cases. This alerts us that the purely input-space-based scores are actually under pressure.
For \texttt{KL} and \texttt{MSS}, the differences coming from jet types are reduced w.r.t. the other scores, while the mass correlation is stronger. (This can be better observed from App. \ref{app:rocs}.)
As shown in the previous section, the KL divergence is strongly correlated with jet mass, which coincides with this observation. Overall, \texttt{KL} slightly outperforms the others in most of the cases.
Directly built on the raw input features, \texttt{MSS} competes with other deep learning based scores and surpasses the other two input-space-based scores in several test sets. This again reminds us that the reconstruction based anomaly detection might be missing something critical.
%This imposes several questions worth thinking about: \textcolor{red}{are we learning more critical information using DNNs?}

Although we see the VAE model is able to give reasonable AUCs for the test sets, there are cases where almost all the metrics lose their competence. Since the mass correlation is obvious, low-mass jets generally give low AUCs. This brings difficulties in tagging jets with lower masses.
For instance, $W$ jets with a mass of 80 GeV can only reach the AUC $\sim$ 0.68 in the best case. We will see how this can be amended in the next section.

%\subsection{Failure Analysis}

%More details  please see App. \ref{app:ood}.

%We have seen that the naively trained VAE got mass correlation and thus fails at tagging jets at lower mass.
%Actually, this failure is not unique in jet tagging problems. As known in deep learning community, density estimation based deep generative models in anomaly detection are generally encountered with the problem of assigning higher likelihood to OoD samples with simpler and less pixel inputs. \cite{2018arXiv181009136N} %add other refs
% Robustscaling + simple gaussian modeling gives already very good estimations in low-level features
%Another interesting thing is, according to our experiments, a simple standard gaussian density function with appropriate scaling of features suffices at anomalous jet tagging tasks.  

%%%%%%%%%%%%%%%%%%%%%%%%%%%%%%%%%%%%%%%%%%%%%%%%%%%%%%%%%%%%%%
\subsection{\label{sec:mass_deco}Mass Decorrelation -- DisCo-VAE}

\begin{figure}[htb!]
    \centering
    \includegraphics[width=0.5\textwidth]{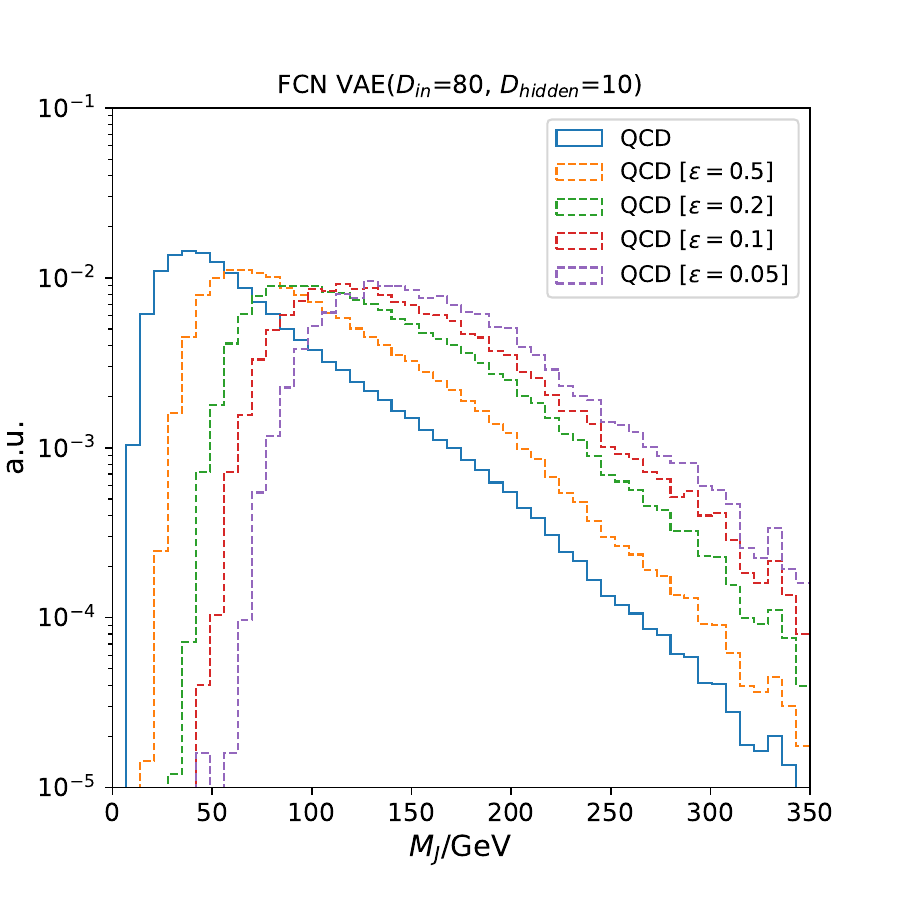}
    \caption{QCD background mass distributions after thresholding VAE reconstruction errors at different background efficiencies $\epsilon$.}
    \label{fig:m_sculpting}
\end{figure}

MSE-based autoencoders, be it built with four vectors or in the format of images, are highly correlated with the jet mass, when tagging anti-QCD anomalies \cite{Heimel:2018mkt}. And we see the mass correlation persists for all the anomaly scores investigated previously. In Fig. \ref{fig:m_sculpting}, we show the mass-sculpting effects of the reconstruction error based tagger. Higher reconstruction error selects a sample of high-mass QCD jets. 

Mass decorrelation is an important topic in general search for resonances using large-radius jets. For example, the bump-hunting analysis utilizes orthogonal information w.r.t. jet mass to reduce the background and then carries out a bump search in the mass dimension.
%, such as bump-hunt \cite{Collins:2018epr}. 
So we dedicated a study to mass-decorrelated taggers in this part. Previous works have utilized adversarial training to decorrelate jet mass for either classifiers \cite{ATL-PHYS-PUB-2018-014, Bradshaw:2019ipy, Shimmin:2017mfk, Louppe:2016ylz} or autoencoders \cite{Heimel:2018mkt}. However, adversarial training is difficult to tune and takes much more computational resources. We instead employ the distance correlation (DisCo) regularization \cite{Kasieczka:2020yyl} as a mass-decorrelation baseline. 

%\paragraph{DisCo-VAE}
In the DisCo approach, a distance correlation regularization term is added to the VAE loss as indicated in Eq. \ref{eqn:disco-vae}:

\begin{equation}
\label{eqn:disco-vae}
    \mathcal{L}_{\rm DisCo-VAE} = \mathcal{L}_{\rm \beta-VAE} + \kappa R_{\rm DisCo} \, ,
\end{equation}
where the regularizer $R_{\rm DisCo}$ is defined in Eq. \ref{eqn:r_disco} as the distance correlation between the VAE loss and the jet mass. 
\begin{equation}
\label{eqn:r_disco}
    R_{\rm DisCo} = {\rm dCor}(M_J,\mathcal{L}_{\rm \beta-VAE})
\end{equation}
Distance correlation \cite{2008arXiv0803.4101S} is a measure of the non-linear correlation between two random variables. Independent variables will give a distance correlation of 0. The distance correlation ${\rm dCor}(X,Y)$ between variable $X$ and $Y$  is defined as in Eq. \ref{eqn:dcor}, with the distance covariance ${\rm dCov}(X,Y)$ defined in Eq. \ref{eqn:dcov}, where ($X'$, $Y'$) and ($X''$, $Y''$) are independent and identically distributed samples of ($X$, $Y$). ${\rm dVar} (X)$ and ${\rm dVar} (Y)$ are simply the distance variances defined by ${\rm dVar}(\cdot) = {\rm dCov}(\cdot, \cdot)$. And $\textrm{cov}(\cdot, \cdot)$ in Eq. \ref{eqn:dcov} denotes the classical Pearson covariance.

% the definitions here slightly differs from the original Distance Correlation notations. Here we use dCor to denote the squared distance correlation, and use dCov to denote the squared distance covariance in the reference.

\begin{equation}
\label{eqn:dcor}
    {\rm dCor}(X,Y) = \frac{{\rm dCov}(X,Y)}{\sqrt{{\rm dVar}(X) {\rm dVar}(Y)}}
\end{equation}

% [2022-03-01] delete the exponent (2) of dCov for consistency
\begin{equation}
\label{eqn:dcov}
  \begin{aligned}
     {\rm dCov}(X,Y) &= \langle \Vert X - X'\Vert \Vert Y - Y'\Vert \rangle
    + \langle \Vert X - X'\Vert \rangle \langle \Vert Y - Y'\Vert \rangle
    -2 \langle \Vert X - X' \Vert \Vert Y - Y'' \Vert \rangle  \\
    &= {\rm cov}(\Vert X - X' \Vert, \Vert Y - Y' \Vert ) 
    - 2 {\rm cov}(\Vert X - X' \Vert, \Vert Y - Y'' \Vert )
  %       {\rm dCov}^2(X,Y) &= \langle |X - X'| |Y - Y'| \rangle
  %  + \langle |X - X'| \rangle \langle |Y - Y'| \rangle
  %  -2 \langle |X - X'| |Y - Y''| \rangle  \\
  %  &= {\rm cov}(\Vert X - X' \Vert, \Vert Y - Y' \Vert ) 
  %  - 2 {\rm cov}(\Vert X - X' \Vert, \Vert Y - Y'' \Vert )
  \end{aligned}
\end{equation}

% mass deco results
For DisCo-VAE, we tested different $\kappa$ values ($\kappa = 0.1, ~0.5, ~1, ~10, ~100, ~200, ~500, ~1000$), and found that $\kappa \gtrsim 100$ starts to give reasonable mass decorrelation. We trained on the same QCD dataset as used in $\beta$-VAEs, with the same batch size of 100.
To achieve better fitting and make sure both terms in the DisCo-VAE loss (Eq. \ref{eqn:disco-vae}) are optimized, we employed annealing training. We first train with only the $\beta$-VAE loss for 10 epochs, then slowly increase the weight of the DisCo regularizer until it reaches the target value in 10 epochs, and eventually train the full DisCo-VAE objective for another 10 epochs before resetting the regularizer weight back to 0. We repeat this process for 2 cycles, after which we continue training the full objective until the 100th epoch. We use Adam as the optimizer with an initial learning rate of 1e-3 which is changed to 1e-4 after 30 epochs.

\begin{figure}[htb!]
    \centering
    % normal scale
    %\includegraphics[width=0.45\textwidth]{figs/disco_mdeco_kappa100_revised.png}
    %\includegraphics[width=0.45\textwidth]{figs/disco_mdeco_kappa1000_revised.png}
    % log scale
    \includegraphics[width=0.45\textwidth]{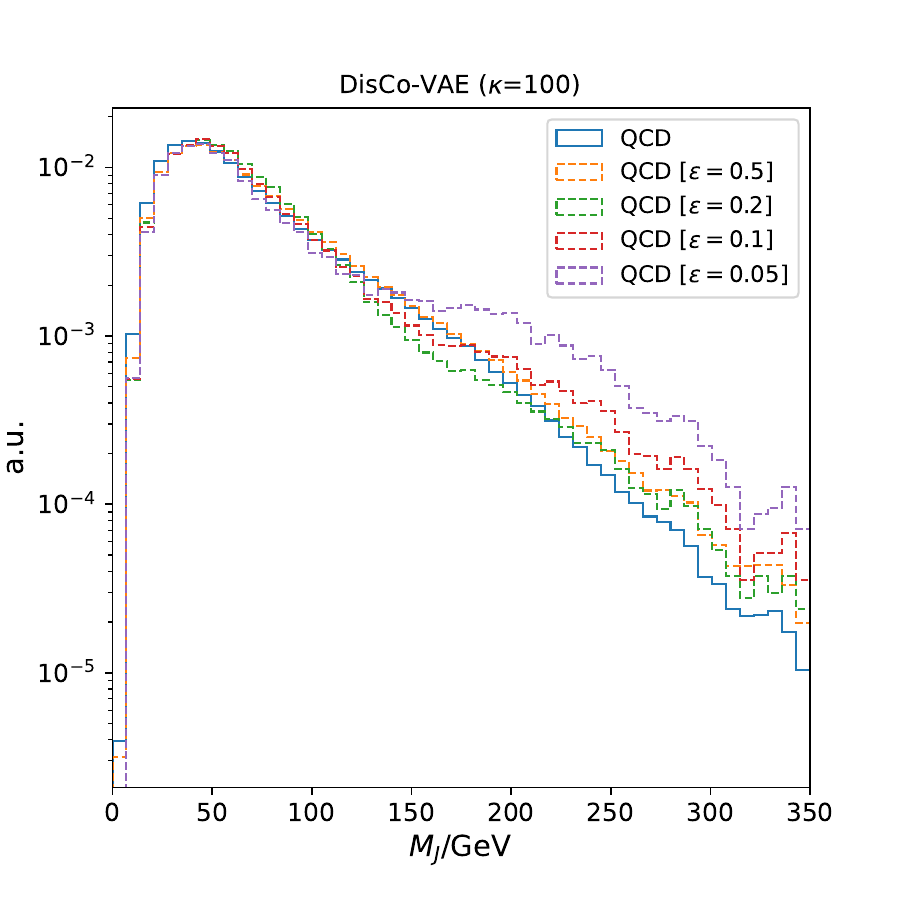}
    \includegraphics[width=0.45\textwidth]{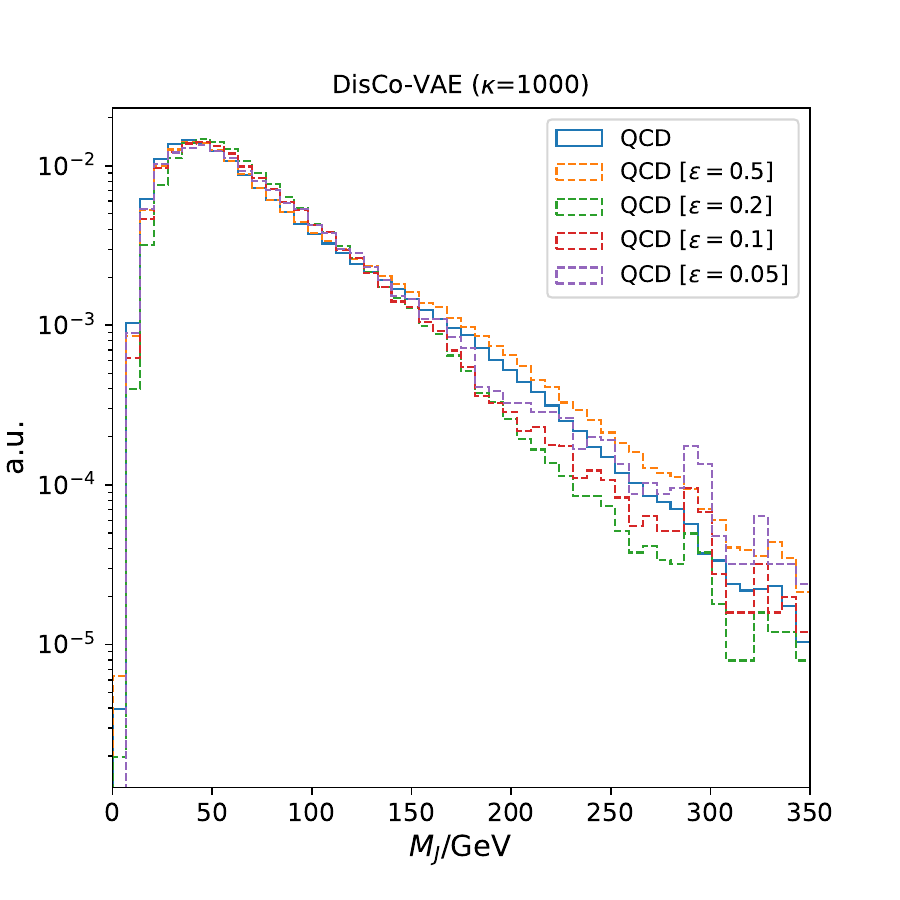}
    \caption{Mass decorrelation effects for DisCo-VAEs with $\kappa=100$ (\textit{Left}) and $\kappa=1000$ (\textit{Right}).}
    \label{fig:disco_mdeco}
\end{figure}

\begin{figure}[htb!]
    \centering
    \includegraphics[width=0.45\textwidth]{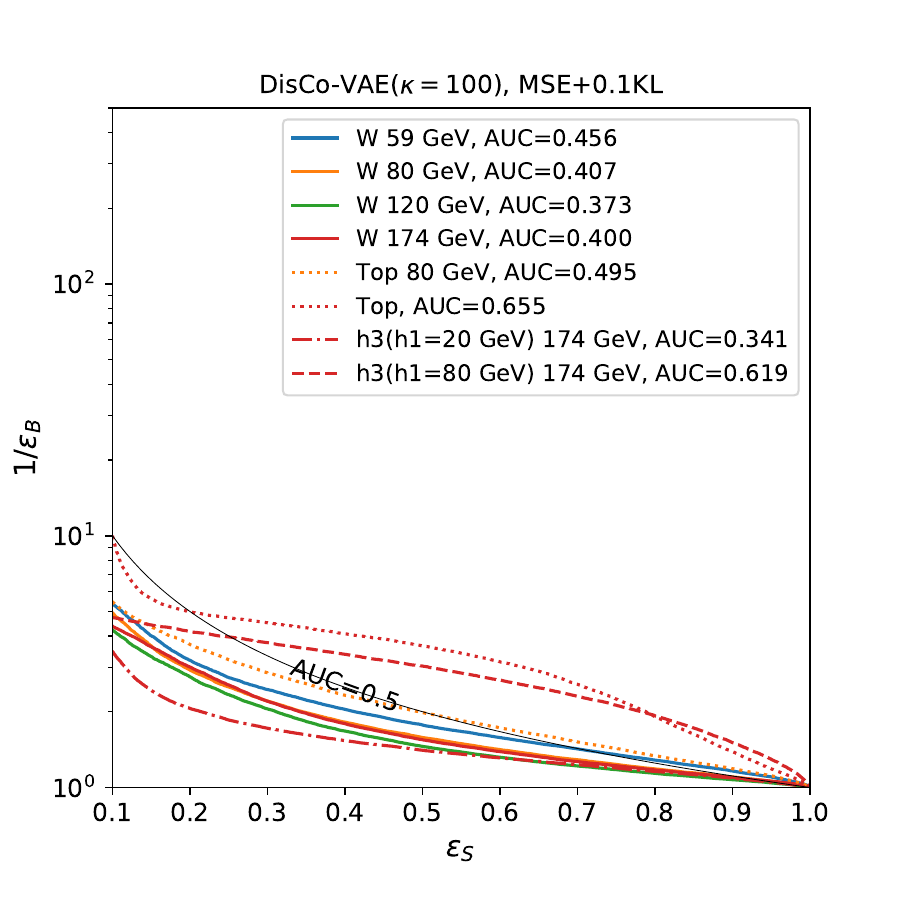}
    \includegraphics[width=0.45\textwidth]{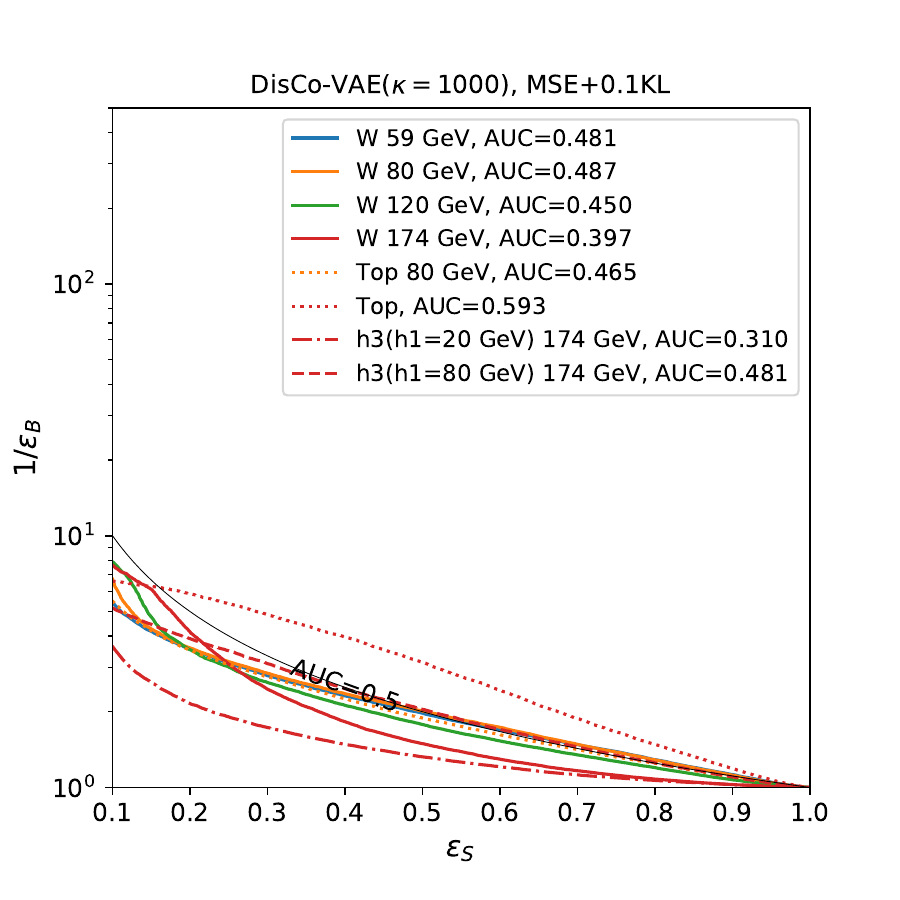}
    \caption{ROC curves for DisCo-VAE with $\kappa=100$ (\textit{Left}) and $\kappa=1000$ (\textit{Right}).}
    \label{fig:disco_rocs}
\end{figure}

In Fig. \ref{fig:disco_mdeco}, we show mass decorrelation effects for DisCo-VAE with $\kappa=100$ and $\kappa=1000$ respectively.
To see the performance of DisCo-VAEs in jet tagging, we employ the $\beta$-loss (MSE+$\beta$KL) which is decorrelated with the jet mass as the anomaly score here. ROC curves for all the test sets are presented in Fig. \ref{fig:disco_rocs}.
Although $\kappa=1000$ gives very good mass decorrelation, we find that it's not able to preserve enough anomaly detection capability.
%Summary AUCs are recorded in Table \ref{tab:aucs_mass_deco}.
For all the test sets, poor anomaly detection performance is observed compared with the results presented for non-decorrelated VAEs.
%in Fig. \ref{fig:vae_rocs_elbo} and \ref{fig:vae_rocs_other_metrics}.
There are even a few test sets giving AUCs less than 0.5, which means that the model is even assigning higher likelihood to out-of-distribution samples than in-distribution samples. This reminds us of the failure case discussed in previous discussion. 

%%%%%%%%%%%%%%%%%%%%%%%%%%%%%%%%%%%%%%%%%%%%%%%%%%%%%%%%%%%%%%
\section{\label{sec:oe} Semi-Supervision -- OE-VAE}

Summarizing the observations presented in the previous sections, the current situation is that the mass-correlation is strong for simple VAEs while the mass-decorrelated DisCo-VAE taggers have poor discrimination performance in the full test spectrum. There are cases that OoD samples are even assigned higher probability than InD samples. This calls for novel approaches to the problem under investigation.
When unsupervised learning might find bad minima for which not much discriminating power is enabled, semi-supervised learning may help with this situation.
%\mynote{some review: Semi-supervised learning %\cite{DBLP:journals/corr/KingmaRMW14}}
One approach which might help with gaining higher sensitivity in anomaly detection is Outlier Exposure (OE) \cite{2018arXiv181204606H}. We thus introduce Outlier Exposed VAE (OE-VAE) as a solution.
By injecting some OoD samples in the training process and requiring the model to separate the exposed OoD samples from the training samples, the VAEs will learn problem-oriented heuristics to help detect general OoD anomalies. This approach is similar to jointly training an auxiliary-classifier, in a pseudo-binary manner for which inliers and outliers are required to be separable.   
%This somehow is similar to combining classification with the representation learning process.

% kill two birds with one stone
In this tailored OE-VAE approach, we try to achieve two goals at the same time: \emph{increasing sensitivity to out-of-distribution samples, and decorrelating the jet mass from the anomaly score.} By simply exposing outliers to the VAE training, we hope to obtain a general sensitivity increase by this auxiliary task, not only to the exposed samples but also generalized to other types of anomalies. 
Another advantage of having an outlier dataset is that we thus have a handle to provide extra guidelines leveraged through the outlier samples.
A simple and useful practice is that we match the mass distribution of inlier samples and outlier samples to decorrelate the jet mass. When asking the VAE to separate the outliers from the training set, it guides and filters the information learned by the model. By matching the mass distribution in this process, information contained in the jet mass that was used to help discriminate against signals is excluded from the learned representations. That eventually leads to the mass decorrelation we are interested in.
%The information picked up by the VAE will maximally exclude jet mass since it will not provide discriminative information. 

% mass decorrelation
We employ mass-rescaled boosted $W$ jet samples as outliers exposed to the VAE training process. As mentioned in Section \ref{sec:anomaly_detection}, a spectrum of boosted $W$ jets are produced by rescaling the Standard Model $W$ mass and generated with the process $ pp \to W' \to W(jj) Z(\nu \nu)$. 
%We thus re-weight the mixed W samples to match QCD jet mass distribution
We resampled the mixture of these mass-rescaled $W$ jets to match their mass distribution to that of the QCD samples. 
%The reweighting is achieved using a deep neural network with fully connected layers.
This procedure is similar to the classical approach ``planing'' \cite{ATL-PHYS-PUB-2018-014} in principle. One may of course use other modern techniques to decorrelate mass. However, in the setting of Outlier Exposure, it comes natural to take advantage of the outlier dataset.

The learning objective of OE can be written as adding a penalty term contributed from outliers $x' \in \mathcal{X}_{\rm OoD}$ to the original VAE loss term, as in Eq. \ref{eqn:loss_oe}:
\begin{equation}
\label{eqn:loss_oe}
   \mathbb{E}_{(x \sim \mathcal{X}_{InD})} \mathcal{L}_{\rm \beta-VAE}(x) + \lambda \mathbb{E}_{(x \sim \mathcal{X}_{InD}, \, x' \sim \mathcal{X}_{OoD})} \mathcal{L}_{\rm OE} (x, x'), \,
     %   \mathcal{L} (\theta, \phi) = \mathbb{E}_{(x \sim D_{InD})} \mathcal{L}_{\rm \beta-VAE}(x) + \lambda \mathbb{E}_{(x \sim D_{InD}, \, x' \sim D_{OoD})} \mathcal{L}_{\rm OE} (x, x'), \,
\end{equation}
where $\lambda$ controls the relative strength of OE. The OE loss term $\mathcal{L}_{\rm OE}$ gains its concrete form according to tasks at hand. 
%\mynote{Stochastic gradient descent, effectively reduces means of samples into ..}
For convenience, we rewrite the OE-VAE loss function as $\mathcal{L}_{\rm OE-VAE} = \mathcal{L}_{\beta-\textrm{VAE}} - \lambda \mathcal{L}_{\rm OE}$. Thus in the training process, we try to maximize $\mathcal{L}_{\rm OE}$ as to minimize $\mathcal{L}_{\rm OE-VAE}$. The outlier exposure can be performed either in input space or in latent space. In the following we discuss both cases.

\begin{itemize}
    \item MSE-OE: In the input space, the OE loss can be written in the sigmoid activation of the difference between mean squared reconstruction error of InD samples and OoD samples. This is very similar to an auxiliary task of classification between InD and OoD:
    \begin{equation}
    \mathcal{L}_{\rm OE} = \sigma_{\rm sigmoid} (\textrm{MSE}_{\rm OoD} (x', \hat x') - \textrm{MSE}_{\rm InD} (x, \hat x))
    \end{equation}
    \item KL-OE: 
    %\alarm{KL leading order $\sim z^2$; while MSE leading order $\sim x^2$}
    %the data distribution in the dimension-reduced latent space is regularized and densely covered. 
    %we assume that the geometry would be flattened and more linearized in the low-dimensional latent space. 
    %\alarm{Mapped from the high-dimensional input space to the low-dimensional latent space which captures the intrinsic dimensions of the data distribution, the latent space is a vector space densely covered by the data points. }
    The Euclidean latent space is assumed to capture the intrinsic dimensions of the original non-linear data manifold in the high-dimensional input space. And the distance in the latent space is more directly linked with the concept of jet ``similarity''.    
    When we get the KL divergence as the metric for latent representation similarity, we can impose the margin loss to express the relative distance between InD samples and OoD samples. We employ the following loss function for the latent space:
    \begin{equation}
    \mathcal{L}_{\rm OE} = \min \{0 , \mathcal{L}_{\rm KL, OoD}(z') - \mathcal{L}_{\rm KL, InD}(z) - \texttt{margin}\}, ~\,
    \end{equation}
    with $z$ and $z'$ denoting latent representations of inlier samples and outlier samples respectively.
    This will encourage outlier samples to have a larger KL divergence above a specific margin.
\end{itemize}
Of course these are not unique choices for the OE loss. But they have been effective in our studies.

\paragraph{Training Setup}
We exposed 120,000 OoD samples \footnote{Regarding the outlier sample size, we have tried different OE sizes ranging from 4/5 to 1/10 of the QCD training set size. And we didn't find significant performance difference due to the sample size. The only issue one should be aware of is that when the OE sample size is too small ($\textrm{training set size} / \textrm{OE sample size} \leq 6$), the model might overfit early but by checkpointing one can ameliorate this issue properly.} consisting of boosted (mass-rescaled) $W$ jets, which are resampled to match the mass distribution of the QCD background.
%$\lambda$ = 50, 100, 200, 500, 750, 1000. 
Increasing $\lambda$ puts higher weight on the supervision part and, at the same time,
enforces stronger mass-decorrelation. Thus $\lambda=500$ is chosen for MSE-OE accordingly to make sure both the VAE loss and the OE penalty are equally optimized.
For KL-OE, we have $\lambda=2$ and \texttt{margin} set to 1 according to mean values of the KL divergence. The models are trained for 50 epochs with a batch size of 100. As in the DisCo-VAE training,
we cyclically anneal $\lambda$ \footnote{Starting from the 10th epoch, we slowly increased $\lambda$ from 0 to the target value in 5 epochs, followed by another 5 epochs' training on the full objective, then started over this process for 3 times. After, the model was trained with the full objective until the 50th epoch.} in the training process to achieve a better balance between the optimization of the VAE loss and the OE loss. Adam with a learning rate of 1e-3 is used for optimization.
%The OE training can be carried out either from scratch or in the manner of fine-tuning a pre-trained VAE model. We report results for the models that were trained from scratch in the following. 
%\begin{figure}
%    \centering
%    \includegraphics[width=0.5\textwidth]{figs/oe_mdeco_predist.png}
%    \caption{Caption}
%    \label{fig:m_dist}
%\end{figure}

\paragraph{Results}

\begin{figure}
    \centering
    \includegraphics[width=0.45\textwidth]{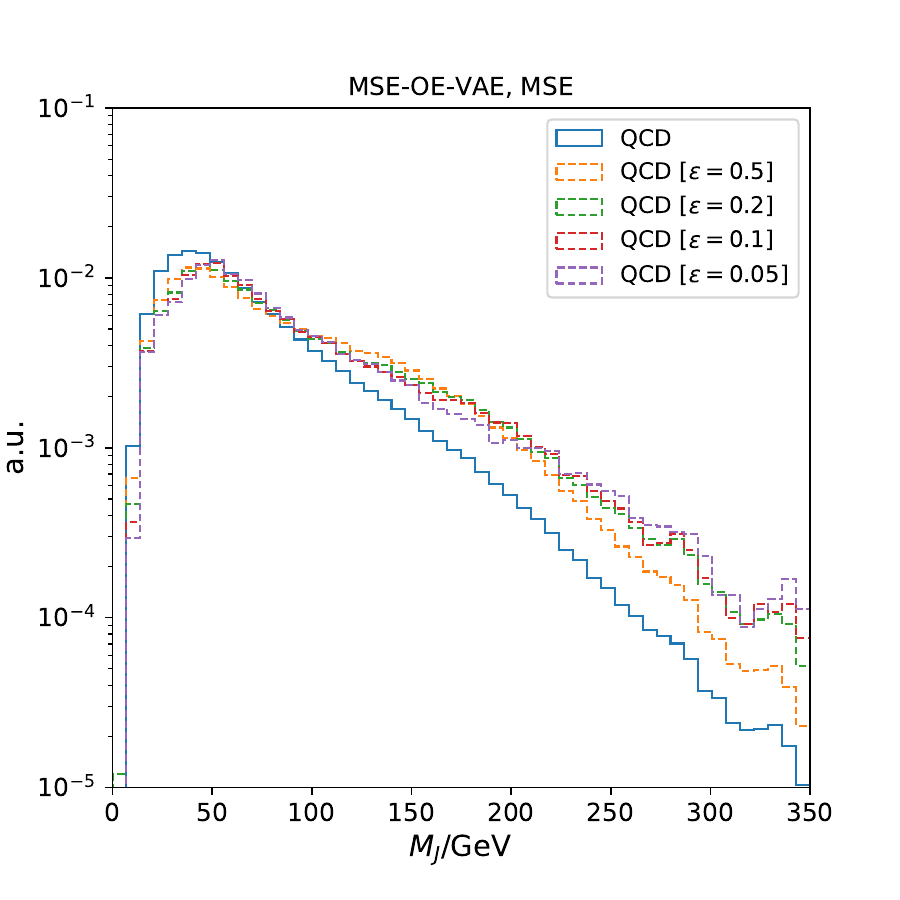}
    \includegraphics[width=0.45\textwidth]{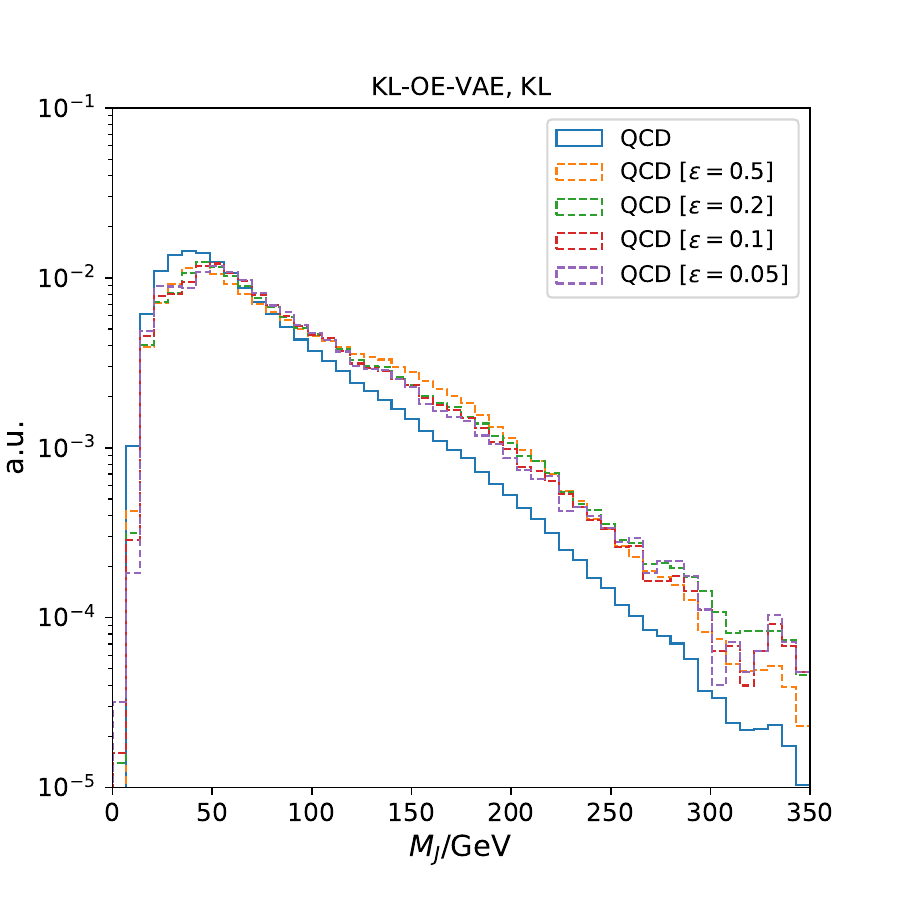}
    \caption{Mass decorrelation effects for MSE-OE-VAE (\textit{Left}) and KL-OE-VAE (\textit{Right}).}
    \label{fig:mdeco_oe}
\end{figure}

We first examine the mass decorrelation effects for OE-VAEs. The results are shown in Fig. \ref{fig:mdeco_oe} for MSE-OE-VAE and KL-OE-VAE respectively. Excellent mass decorrelation effects are achieved in both scenarios.
One thing to keep in mind is that only outlier exposed metrics can be used for mass-decorrelated taggers. The equivalent ``planing'' effects only affects one space: the input space or the latent space. One should match the anomaly metric with the OE training scenario, i.e., if one trains OE-VAE in the input space, then accordingly using the input space MSE as the anomaly score will provide the desired mass decorrelation.

\begin{figure}
    \centering
    \includegraphics[width=0.45\textwidth]{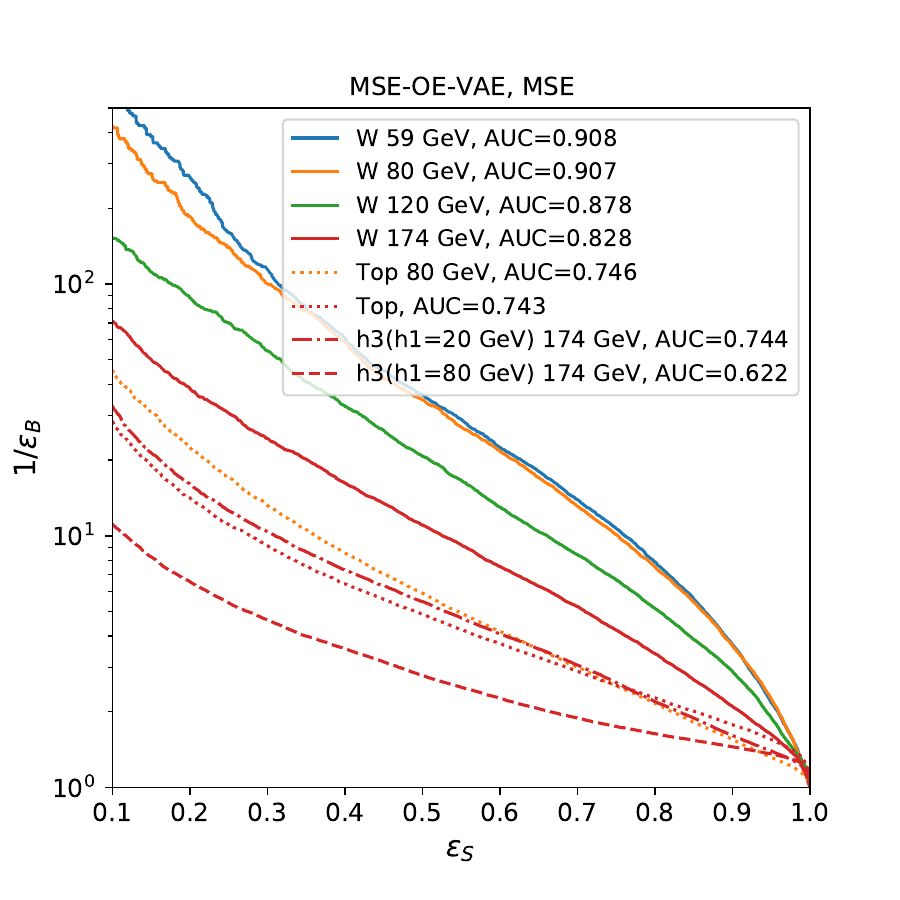}
    \includegraphics[width=0.45\textwidth]{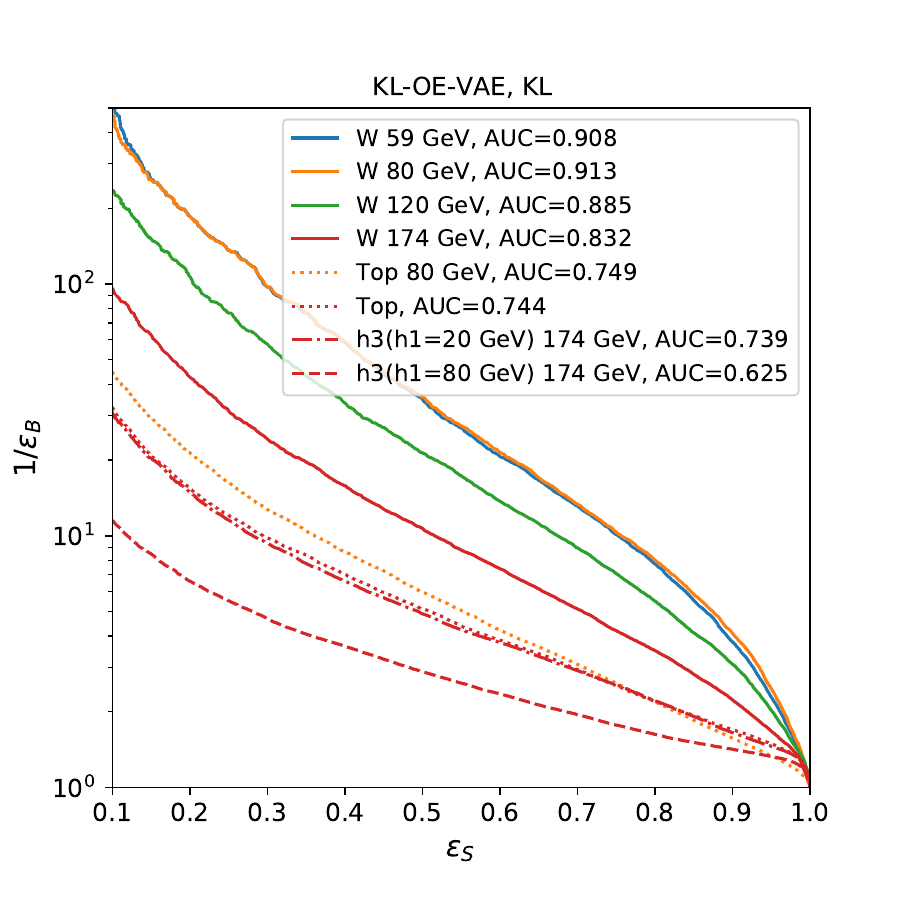}
    \caption{\textbf{Left}: ROC curves for MSE-OE-VAE ($\lambda=500$); \textbf{Right}: ROC curves for KL-OE-VAE ($\lambda=2, ~ \texttt{margin}=1$).}
    \label{fig:rocs_oe}
\end{figure}

\begin{figure}
    \centering
    \includegraphics[width=0.9\textwidth]{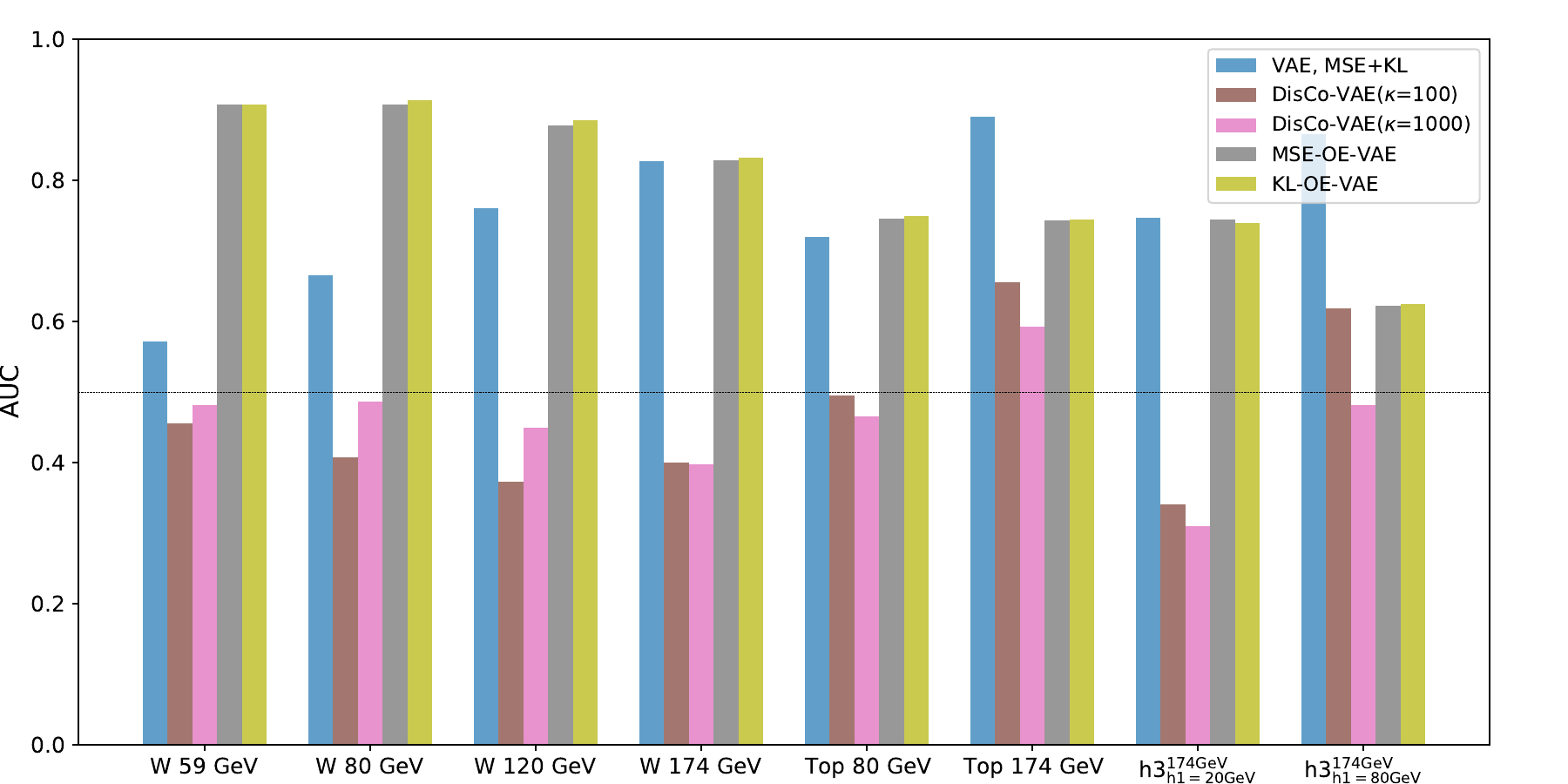}
    \caption{Summary plot of AUCs for mass-decorrelated models, in the complete spectrum of test signal samples. The simple VAE with the anomaly metric of NLL from the previous section is kept as a reference to models without mass decorrelation.}
    \label{fig:aucs_mdecomodels}
\end{figure}

When presenting anomaly detection performance of OE-VAEs, we accordingly employ the mass-decorrelated anomaly scores, i.e. mean squared reconstruction error for MSE-OE-VAE and latent KL divergence for KL-OE-VAE.
In Fig. \ref{fig:rocs_oe}, we present ROC curves for all the test signal samples for MSE-OE-VAE and KL-OE-VAE respectively. In Fig. \ref{fig:aucs_mdecomodels} we plot the AUCs of all the mass-decorrelated models for comparison, with the non-decorrelated VAE as a reference. A summary of AUC numbers is recorded in Table \ref{tab:aucs_mass_deco}.

Both MSE-OE-VAE and KL-OE-VAE give similar improved performance. The first observation is that the $W$ tagging performance is immediately improved. Respectively, a few test sets such as top have lower AUCs compared with VAE. This is mainly due to the extra mass decorrelation since mass decorrelation generally drops the AUCs of high mass jets.
Comparing within mass-decorrelated models, the semi-supervised OE scenario gains better discrimination for top jets than the baseline DisCo-VAEs.
Solely comparing the AUCs in the full spectrum, OE-VAEs outperform DisCo-VAEs in all the test sets. 
%For OE-VAEs, there are two goals at the same time: increasing sensitivity to OoD samples and decorrelating jet mass from the anomaly score. 
There are two efforts within OE-VAEs: the first is to induce more informative representation learning using exposed outliers; the second is to subtract jet mass from the discriminative information. If we gain better anomaly detection performance while decorrelating jet mass, that means the model is learning more discriminative information orthogonal to jet mass than the baseline DisCo-VAEs. If the induced extra discriminative information exceeds the information within the jet mass, we get even higher AUCs than the naive VAEs. That's exactly the case for $W$ jets: the AUCs of OE-VAEs are even higher than that of non-decorrelated VAEs. \footnote{i.e., signals similar to the outliers exposed will gain even further sensitivity improvement despite the jet mass is decorrelated from the anomaly score.}
 Since we employed directly $W$ as outliers exposed, the learned model is most sensitive to it than other test signals. 

%%%%%%%%%%%%%% Summary Table for AUCs %%%%%%%%%%%%%%%%%%%%%%%%
\begin{table}[htb!]
    \centering
    \scalebox{0.9}{
    \begin{tabular}{c|cccccccc}
    \hline 
    \multirow{2}{*}{\backslashbox{Model}{Signal}}  &  \multicolumn{4}{c}{$W$}  & \multicolumn{2}{c}{Top} & \multicolumn{2}{c}{Higgs}\\
    & $W_{59 \textrm{GeV}}$ &  $W_{80 \textrm{GeV}}$ & $W_{120 \textrm{GeV}}$ & $W_{174 \textrm{GeV}}$ & Top$_{80 \textrm{GeV}}$ & Top$_{174 \textrm{GeV}}$ & h3$^{174 \textrm{GeV}}_{\textrm{h1=20GeV}}$ & h3$^{174 \textrm{GeV}}_{\textrm{h1=80GeV}}$\\
    \hline \hline
    DisCo-VAE($\kappa=100$) & 0.456 & 0.407 & 0.373 & 0.400 & 0.495 & 0.655 & 0.341 & 0.619 \\
    % hidden track
    %DisCo-VAE($\kappa=100$, anneal 2 steps) & 0.485 & 0.412 & 0.367 & 0.424 & 0.447 & 0.651 & 0.334   & 0.620 \\
    %DisCo-VAE($\kappa=100$, anneal 2 steps, early stopping (51 epochs)) & 0.504  & 0.460  & 0.418  & 0.448  & 0.484  & 0.675  & 0.383  & 0.627 \\
    %DisCo-VAE($\kappa=100$, early stopping (75 epochs), 3 step annealing) & 0.45129008  & 0.39334744499999996  & 0.3331566425  & 0.358807545  & 0.37083330519206414  & 0.6510614625  & 0.2489292575  & 0.604536175
    %DisCo-VAE($\kappa=100$, anneal 3steps, earlystopping) & 0.530  & 0.475  & 0.411  & 0.438  & 0.453  & 0.689  & 0.340  & 0.654 
    %DisCo-VAE($\kappa=100$, anneal 3steps, epoch=100) &     0.520  & 0.462  & 0.404  & 0.467  & 0.492  & 0.705  & 0.376  & 0.671\\
    DisCo-VAE($\kappa=1000$)& 0.481  & 0.487  & 0.450  & 0.397  & 0.465  & 0.593  & 0.310  & 0.481\\
    %DisCo-VAE($\kappa=1000$, anneal 2 steps)& 0.430 & 0.363 & 0.301 & 0.354 & 0.405 & 0.617 & 0.298   & 0.561 \\
    %DisCo-VAE($\kappa=1000$, anneal 2steps, early stopping) & 
    %DisCo-VAE($\kappa=1000$, anneal 3steps, early stopping) & 0.500  & 0.485  & 0.450  & 0.499  & 0.483  & 0.590  & 0.391  & 0.464 \\
    %DisCo-VAE($\kappa=1000$, anneal 3steps, epoch=100) & 0.541  & 0.523  & 0.446  & 0.441  & 0.449  & 0.597  & 0.387  & 0.499 \\
    MSE-OE-VAE  & 0.908  & 0.907  & 0.878  & 0.828  & 0.746  & 0.743  & 0.744  & 0.622 \\
    % wrong top m80
    %MSE-OE-VAE  & \textbf{0.908} & 0.907 & 0.878 & 0.829 & \textbf{0.729} & 0.743 & \textbf{0.743} & 0.620 \\
    % hidden track
    %MSE-OE-VAE(early stopping, n=20k) & 0.891  & 0.881 & 0.837 & 0.748 & 0.651 & 0.668 & 0.657 & 0.544   \\
    %MSE-OE-VAE(early stopping) & 0.908  & 0.902  & 0.867  & 0.801  & 0.695  & 0.717  & 0.719  & 0.578 \\
    KL-OE-VAE   & 0.908  & 0.913  & 0.885  & 0.832  & 0.749  & 0.744  & 0.739  & 0.625 \\ % KL = 0.00307
    % wrong top m80
    %KL-OE-VAE   & \textbf{0.908} & \textbf{0.913} & \textbf{0.885} & \textbf{0.832} & 0.721 & \textbf{0.744} & 0.736 & \textbf{0.624}\\ 
    %KL-OE-VAE(early stopping) & 0.921  & 0.916  & 0.881  & 0.814  & 0.714  & 0.719  & 0.749  & 0.598 \\
    \hline
    \end{tabular}}
    \caption{AUCs for mass-decorrelated models in the complete spectrum of test signal samples.}
    \label{tab:aucs_mass_deco}
\end{table}
%%%%%%%%%%%%%%%%%%%%%%%%%%%%%%%%%%%%%%%%%%%%%%%%%%%%%%%%%%%%%%

\paragraph{DisCo-VAE v.s. OE-VAE}

%To see more clearly how much performance improvement can be achieved by OE-VAEs, we compare with DisCo-VAE which is also mass-decorrelated.
%[W Jets] even dropped AUROC curves. (OE wins)
%As an extra note: as same as in OE training, mass decorrelation effect is also relevant to which objective is trained with DisCo. For instance, if trained on reconstruction error, the KL divergence will still have some mass dependence. 
To make a more quantitative examination of the mass-decorrelation quality, we employ the measure based on the Jensen-Shannon Divergence (JSD, Eq. \ref{eqn:jsd}), which is the symmetric version of the KL Divergence to measure the similarity between two probability distributions. Generally speaking, the lower JSD is, the better two distributions match. So for an anomalous jet tagger, we expect a low JSD, and at the same time high AUCs.
\begin{equation}
\label{eqn:jsd}
    D_{\rm JS} [p(m) \Vert p'(m)] = \frac{1}{2} (
    D_{\rm KL}[p(m) \Vert \bar p(m)] + 
    D_{\rm KL}[p'(m) \Vert \bar p(m)] ) , \, ~
    \bar p(m) = \frac{p(m) + p'(m)}{2} 
\end{equation}

\begin{figure}
    \centering
    \includegraphics[width=0.6\textwidth]{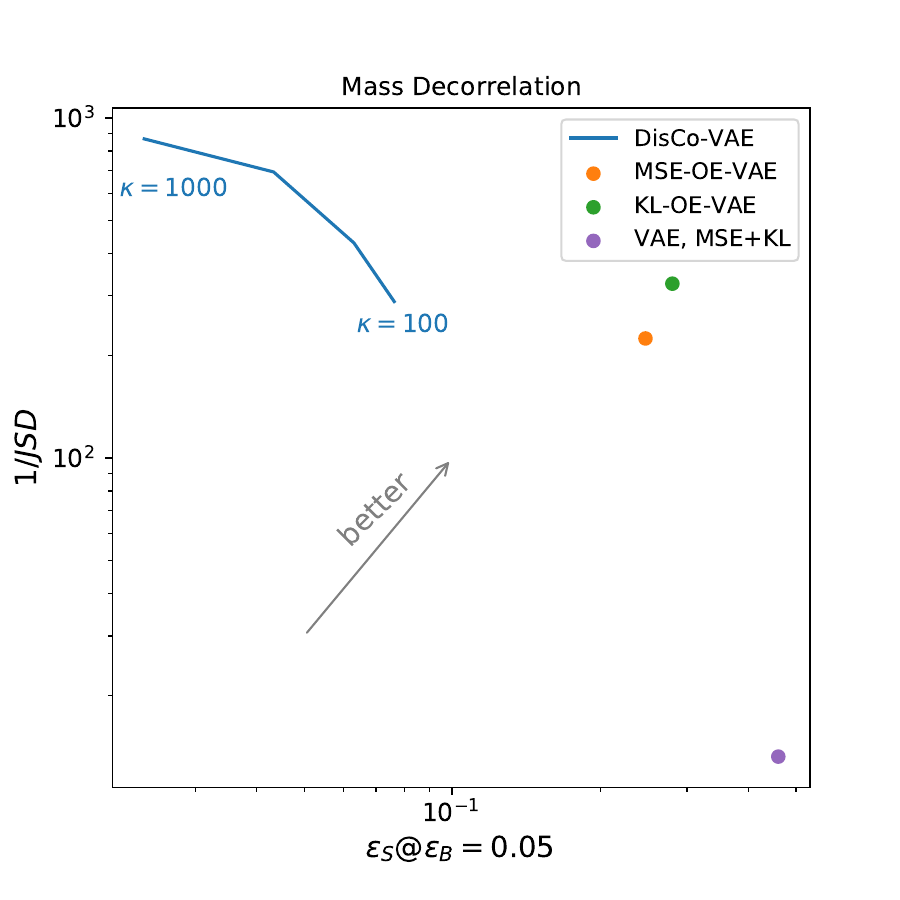}
    \caption{Jensen-Shannon Divergence v.s. Signal Efficiency at the background efficiency of 5\% for mass-decorrelated VAEs. Different $\kappa$ values ($\kappa=$ 100, 200, 500, 1000) are shown for DisCo-VAE. As a reference, we also plot the non-decorrelated VAE with the anomaly metric NLL.}
    \label{fig:jsd}
\end{figure}

In Fig. \ref{fig:jsd}. We plot inverse JSD w.r.t. signal efficiency at the background efficiency of 5\% for the top test set, which is a held-out class when we trained the OE-VAEs with $W$ jets exposed. OE-VAEs generally have much higher signal efficiency w.r.t. DisCo-VAEs at the same mass-decorrelation level. 

%\cite{ATL-PHYS-PUB-2018-014, Bradshaw:2019ipy}

This section is not a full account of the application of Outlier Exposure, or more broadly semi-supervised learning, in the context of generative-model-based anomalous jet tagging. 
We have shown in Section \ref{sec:anomaly_detection} that the naive generative modeling in the unsupervised extreme, with our current settings (training dataset, encoding architecture, generative modeling), is not enough to bring a powerful generic anomalous jet tagger. The standard VAE models fail at detecting several signals. With mass decorrelation implemented, the VAE models almost lose their competence at the full spectrum of test sets. It's promising that OE-VAEs retain very good anomaly detection power while at the same time being mass-decorrelated. We only scratched the surface of semi-supervision and auxiliary tasks for representation learning in this work. This direction is definitely worth more investigation.

%Unsupervised learning without any guidance is extremely vulnerable and brittle.

%%%%%%%%%%%%%%%%%%%%%%%%%%%%%%%%%%%%%%%%%%%%%%%%%%%%%%%%%%%%%%
\section{\label{sec:summary}Summary and Discussions}

Unsupervised learning is a promising approach for model-independent new physics searches at the LHC. We carefully investigated Variational Autoencoders for anti-QCD anomalous jet tagging, totally based on low-level input features. 

To better regularize the latent space and achieve a good balance between jet reconstruction and latent space inference, we cast this work in a generalized VAE setting -- $\beta$-VAE, for which the latent KL divergence is placed with a variable weight. We examined the VAE properties, including reconstruction performance, generation ability, and latent representations, of the trained VAEs. High-level features such as jet $\pt$ and jet mass are correctly reconstructed, and the generative model also performs very well.

To systematically assess the anomalous jet tagging performance, we generated a comprehensive series of test sets including mass-rescaled $W$ jets, top jets, and Higgs jets, as representatives of jet ``prongness". Different anomaly metrics were studied. A strong mass correlation is generally observed for all the anomaly scores studied, leading to smaller and unsatisfying AUCs for low-mass jets. 

As an important application of anti-QCD jet taggers, jet based heavy resonance searches benefit from a mass-decorrelated tagger which makes background estimation accessible. Aiming at a mass-decorrelated tagger, we employed the distance correlation between jet mass and VAE objective as a regularizer. While eliminating mass sculpting effects, the general performance in anti-QCD tagging is rather impractical. This informs us that the generative modeling might not be effective in assigning correct likelihood to data.

To achieve higher sensitivity to out-of-distribution samples and at the same time decorrelate the jet mass from the anomaly score, we employed the Outlier Exposure technique to help learn more tasks-aware latent representations. By injecting some outlier samples to the VAE training process in the manner of auxiliary tasks,
sensitivity to outliers is generally increased. 
By matching the outlier mass distribution to the inlier QCD mass distribution to restrict the information learned by the VAE, we achieved very good mass-decorrelation at the same time. OE-VAEs are compared with the baseline DisCo-VAEs in a two-dimensional metric of the ability of decorrelating jet mass and detecting anomalies.
OE-VAEs are gaining much better performance at the same level of mass decorrelation. We found that Outlier Exposure is a very simple yet effective trick to improve the performance of deep generative models, specifically VAEs in this work, in the scope of anomalous jet tagging. 

% discussions and summary
%{key findings -> secondary findings -> context -> strength & limitations -> what's next -> "so what"}
In summary, we formulate the problem of using generative models (specifically Variational Autoencoders) to detect anomalous non-QCD jets for the LHC. 
Assisted with a comprehensive testing system, we observed that unsupervised learning without any guidelines might not give the optimal solution. Especially, with mass decorrelation the models almost lose the ability to act as an anomaly detector. To solve this problem, a simple semi-supervised approach to enhance the performance and facilitate background estimation is investigated. 
The results show that our method succeeds in decorrelating jet mass and maintaining discriminative power on anomalous samples at the same time. This attempt shows great potential for alternative approaches in addition to pure unsupervised learning.

Despite this effort, there are still several improvements that could be pursued. In this study, only a simple encoding architecture of FCN is used. An LSTM model was also investigated without observing significant improvement, but not much effort is spent on exploring more complex encoding architectures. We expect improved reconstruction ability in that case. In the standard VAE, the latent priors are simply standard multivariate Gaussian distributions. This can be extended to more complex latent priors. Besides these, alternative reconstruction error formats or input space similarity metrics can be explored to better represent the input space structure. And regarding the semi-supervision tasks, the types of outlier samples can affect detection sensitivity according to the discussion in Sec. \ref{sec:oe}. An extensive study in this respect will also be interesting.
We leave these possibilities to future work. 
%(MSE is based on the assumption of factorized Gaussian distribution in input-space.)

\acknowledgments

This work is supported by IVADO Fundamental Research Grant, IVADO Postdoctoral Research Funding, and Natural Sciences and Engineering Research Council of Canada (NSERC). The authors would like to acknowledge Amir Farbin, Debottam Bakshi Gupta, Takuya Nobe, and Johnny Raine for early participation in the project.
TC would like to thank Faruk Ahmed, Aaron Courville and Florian Bordes for helpful discussions on anomaly detection in the general machine learning community.

\appendix

%\section{Comparison with Tilman paper}
%\section{LSTM Models}
%\section{$\beta-$Annealing}

\section{Low-Level Input-Space Features and Jet Images}

%We show that the decoded outputs resemble input distributions.

%In Fig. \ref{fig:generation_features}, we show aggregated distributions of the generated output features (after standardization \footnote{\emph{RobustScaler} in \emph{scikit-learn}
%} of input features) and jet images on the $(\eta, \phi)$ plane.  The generated features $\hat E_i$ approximately follow standard Gaussian distributions as we standardized the input features by removing the median and scaling with the interquartile range before feeding into the VAEs.
%The distributions of the first 10 output energy features $\hat E_i$ are shown in the left panel. 
%\textcolor{red}{removes the median and scales the data according to the quantile range (defaults to IQR: Interquartile Range). The IQR is the range between the 1st quartile (25th quantile) and the 3rd quartile (75th quantile).}

\begin{figure}[htb!]
    \centering
    \includegraphics[width=0.4 \textwidth]{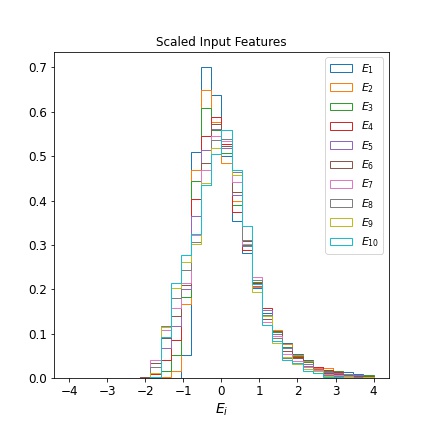}
    \includegraphics[width=0.4 \textwidth]{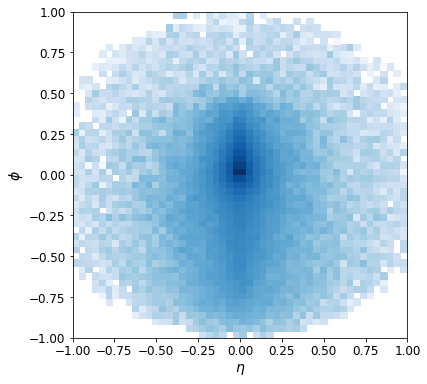}
    \includegraphics[width=0.4 \textwidth]{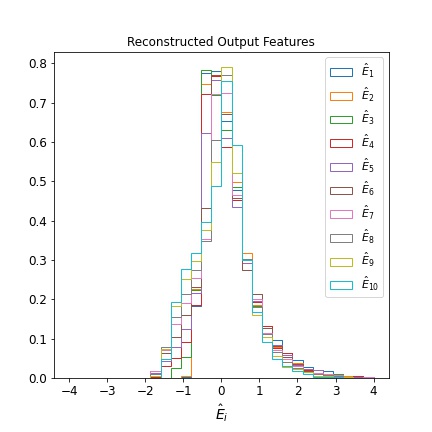}
    \includegraphics[width=0.4 \textwidth]{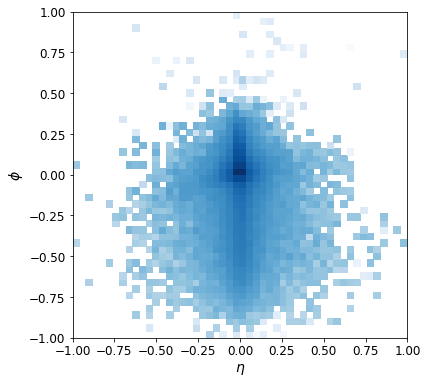}
    \includegraphics[width=0.4 \textwidth]{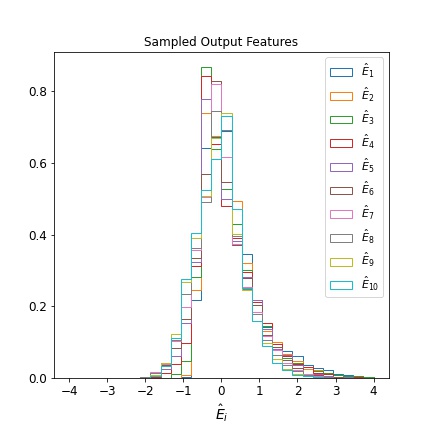}
    \includegraphics[width=0.4 \textwidth]{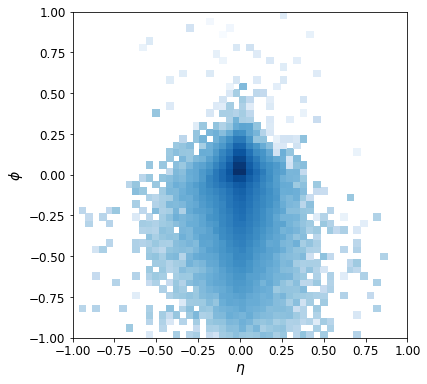}
    \caption{ In the left column, we show the distributions of input-space features for the dimensions of the first 10 constituent energies {$E_i$}. We are using $E_i$ to denote the rescaled input-space features here, rather than the original values of $E_i$.
    In the right column, corresponding jet images are presented.
    \textbf{First row:} input features and input jet images; \textbf{Second row:} reconstructed output features and reconstructed jet images; \textbf{Third row:} VAE generated output features by randomly sampling the latent representations and the generated jet images.
    %In the upper part, we show input feature distributions for the dimensions of the first 10 constituent energies {$E_i$} and input jet images. In the lower part, we show (\textbf{Left}) VAE generated feature distributions for the first 10 energy dimensions {$\hat E_i$} and 
    %(\textbf{Right}) VAE generated jet images on the $(\eta, \phi)$ plane.
    }
    \label{fig:generation_features}
\end{figure}

In Fig. \ref{fig:generation_features}, we show the aggregated distributions of the low-level input-space features and corresponding jet images on the $(\eta, \phi)$ plane. We show respectively the input features, the reconstructed output features, and the generated output features in the upper, middle, and lower row.

\section{Regularization Strength Affects VAEs' Behaviour}
\label{app:betavae}

\begin{figure}[htb!]
    \centering
    \includegraphics[width=0.32\textwidth]{vae_recon_pt.png}
    \includegraphics[width=0.32\textwidth]{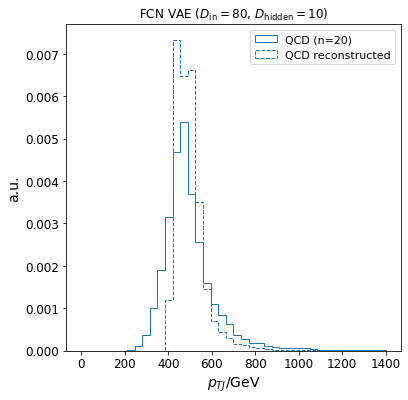}
    \includegraphics[width=0.32\textwidth]{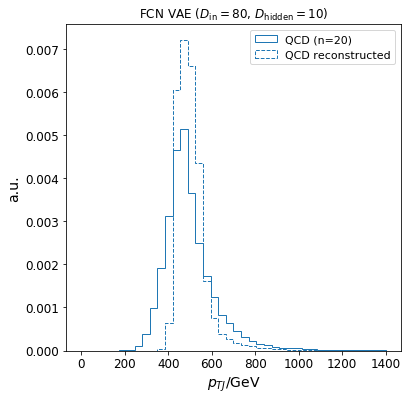}
    \includegraphics[width=0.32\textwidth]{vae_recon_m.png}
    \includegraphics[width=0.32\textwidth]{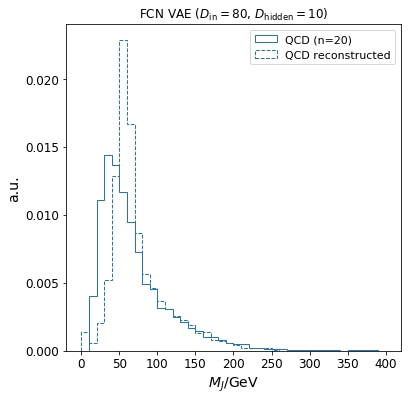}
    \includegraphics[width=0.32\textwidth]{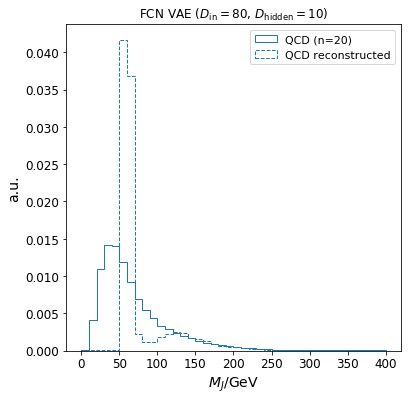}
    \caption{Reconstructed jet $\pt$ (\textit{Upper}) and $M_{J\rm }$ (\textit{Lower}) for different $\beta'$s. \textbf{Left}: $\beta=0.1$; \textbf{Middle}: $\beta=0.5$; \textbf{Right}: $\beta=1.0$.}
    \label{fig:betavae_recons}
\end{figure}

\begin{figure}[htb!]
    \centering
    \includegraphics[width=0.32\textwidth]{vae_latents_tsne_beta0p1.png}
    \includegraphics[width=0.32\textwidth]{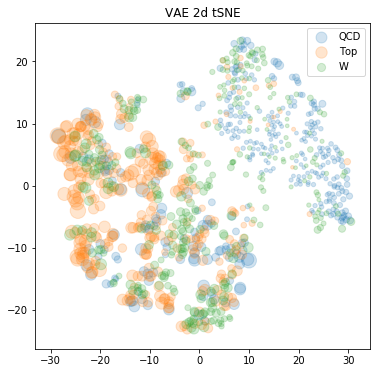}
    \includegraphics[width=0.32\textwidth]{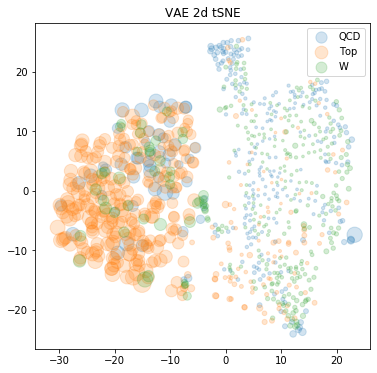}
    \caption{tSNE visualisation of the latent representations (QCD jets (\texttt{Blue}), $W$ jets (\texttt{Green}) and top jets (\texttt{Orange})). \textbf{Left}: $\beta=0.1$; \textbf{Middle}: $\beta=0.5$; \textbf{Right}: $\beta=1.0$.}
    \label{fig:tsne_betas}
\end{figure}

As discussed in Sec. \ref{sec:vae}, the regularization strength $\beta$ will affect the optimization of jet reconstruction and also the latent distribution. In Fig. \ref{fig:betavae_recons}, reconstructed jet observables are shown for different $\beta=0.1, ~0.5, ~1.0$. When $\beta$ increases, reconstruction performance drops. In Fig. \ref{fig:tsne_betas}, tSNE visualisation of latent representations is shown. As $\beta$ increases, in the latent space there is clustering effect emerging. 
%\alarm{a bit more discussion}

%\paragraph{Mass Modes development in large $\beta$}

%\section{Test: training on Top}
%quick test experiments 

\section{Model Comparison}

\begin{figure}[htb!]
    \centering
    \includegraphics[width=0.5\textwidth]{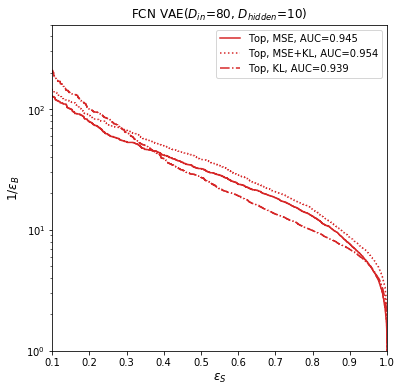}
    \caption{ROC curves of VAE tested on the top tagging reference dataset \cite{Heimel:2018mkt}.}
    \label{fig:comparison_top}
\end{figure}

To facilitate model comparison with previously open-sourced datasets, we show in Fig. \ref{fig:comparison_top} the ROC curves for top samples taken from Ref. \cite{Heimel:2018mkt}. We got AUC = 0.954 (with MSE+KL as the anomaly score) for top jets which is higher than all the AUCs reported there (0.93 for LOLA autoencoder and 0.89 for CNN autoencoder).

\section{Additional ROC Curves for Different Anomaly Scores}
\label{app:rocs}

\begin{figure}[htb!]
    \centering
    \includegraphics[width=0.45\textwidth]{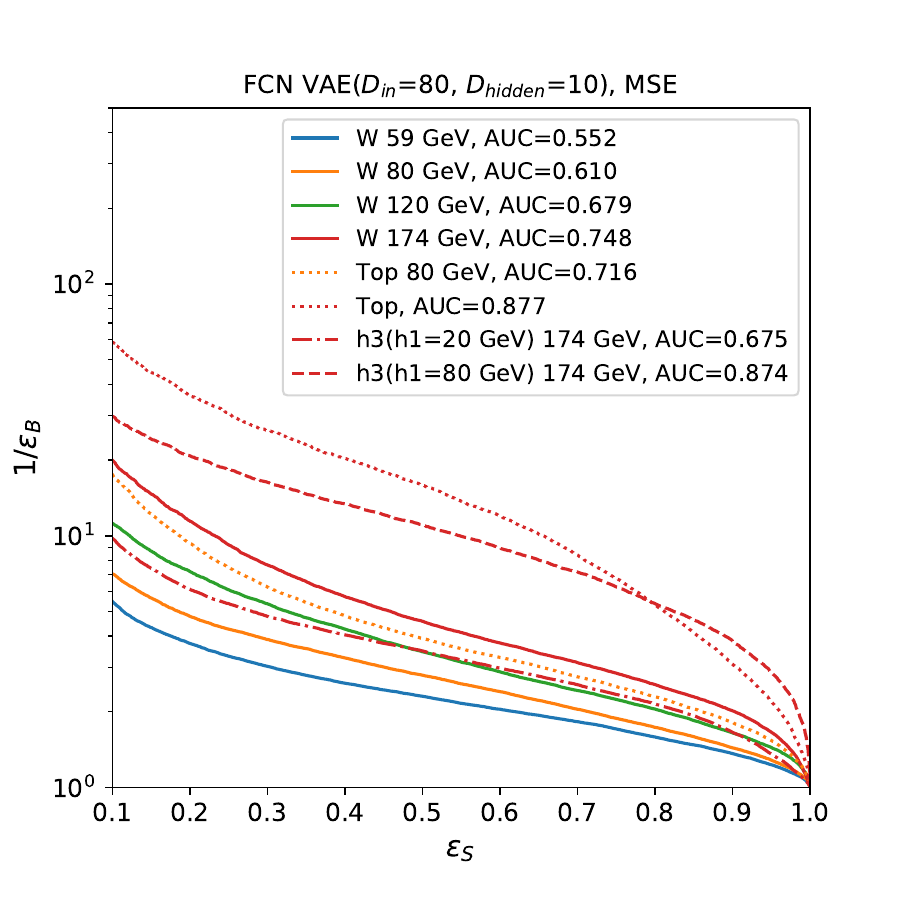}
    \includegraphics[width=0.45\textwidth]{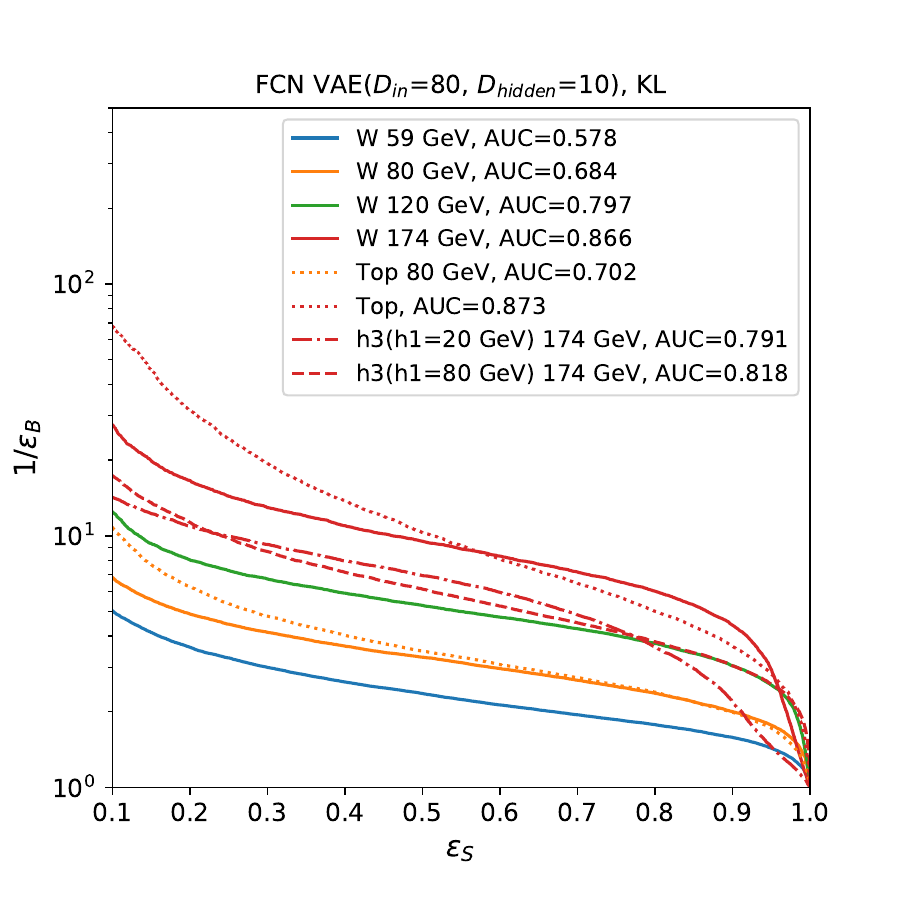}
    \includegraphics[width=0.45\textwidth]{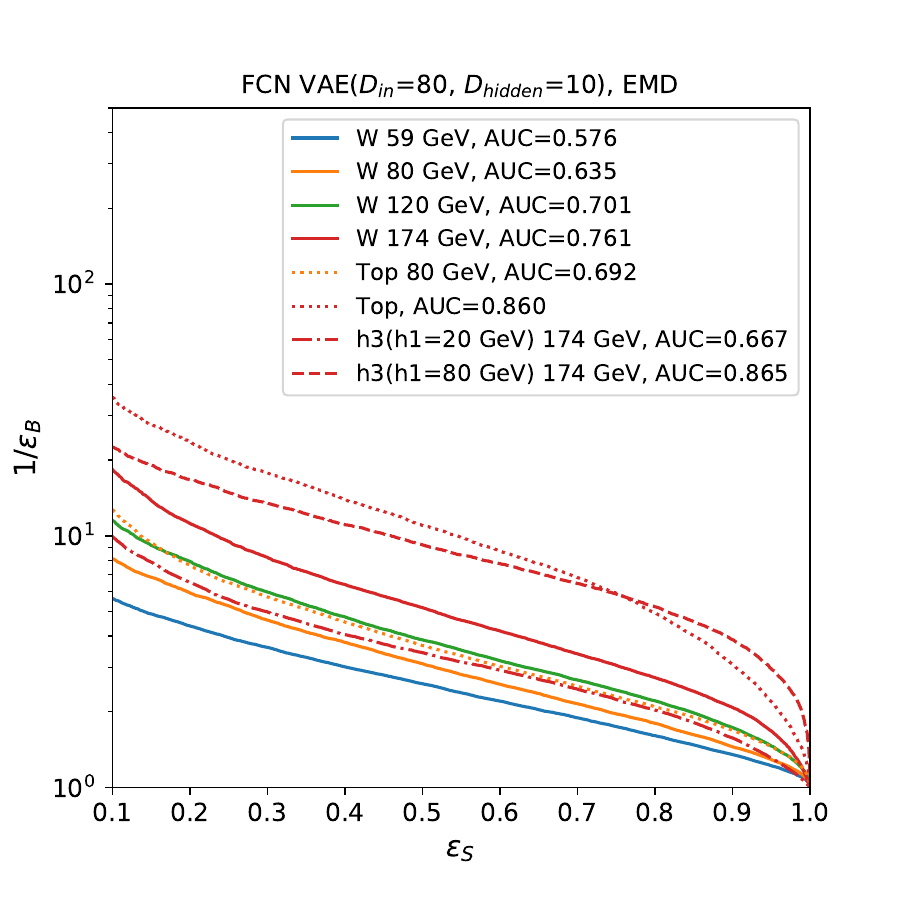}
    \includegraphics[width=0.45\textwidth]{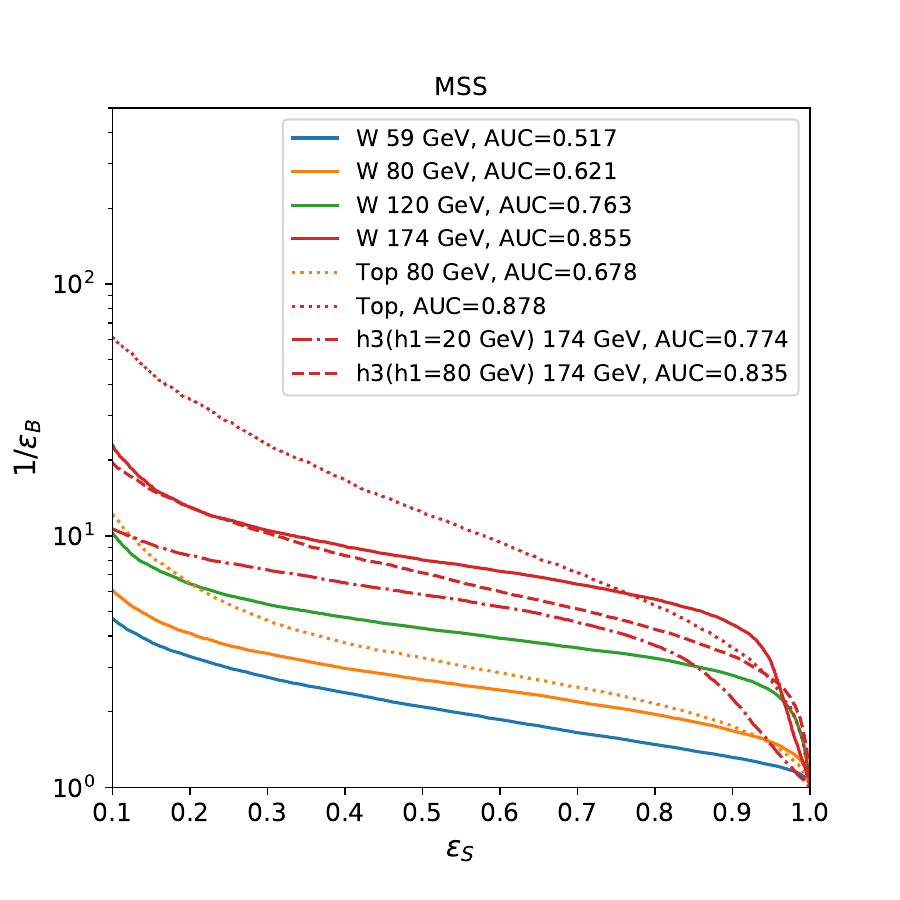} 
    \caption{ROC curves for the spectrum of test signal samples, with the anomaly metric of MSE reconstruction error (\textit{Upper-left}), KL divergence (\textit{Upper-right}), EMD (\textit{Lower-left}) and MSS (\textit{Lower-right}), respectively.}
    \label{fig:vae_rocs_other_metrics}
\end{figure}

In Fig. \ref{fig:vae_rocs_other_metrics}, we present the ROC curves in the complete spectrum of test signal samples, for $\beta$-VAE with the anomaly score of \texttt{MSE}, \texttt{KL}, \texttt{EMD}, and \texttt{MSS}.

\section{Supervised $W$/QCD Classifier v.s. OE-VAE}
\label{app:w/qcd}
\begin{figure}[htb!]
    \centering
    \includegraphics[width=0.6\textwidth]{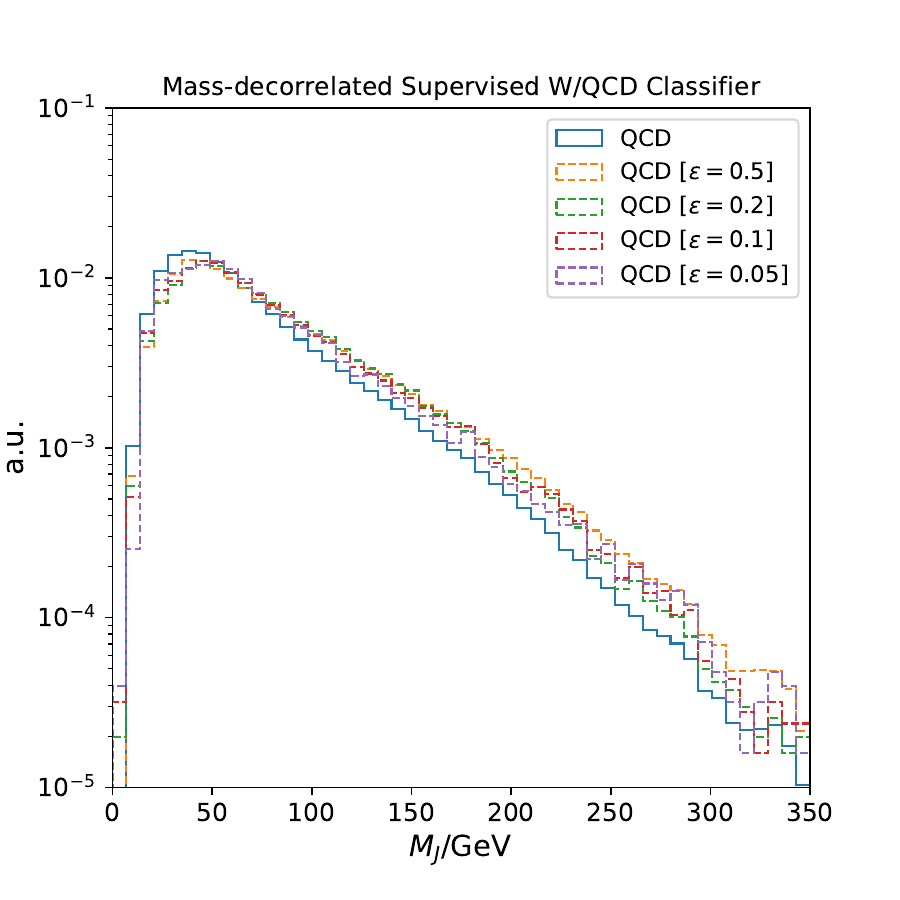}
    \caption{Mass decorrelation effects for the mass-decorrelated supervised $W$/QCD classifier.}
    \label{fig:clf_mdeco}
\end{figure}

\begin{figure}[htb!]
    \centering
    \includegraphics[width=0.45\textwidth]{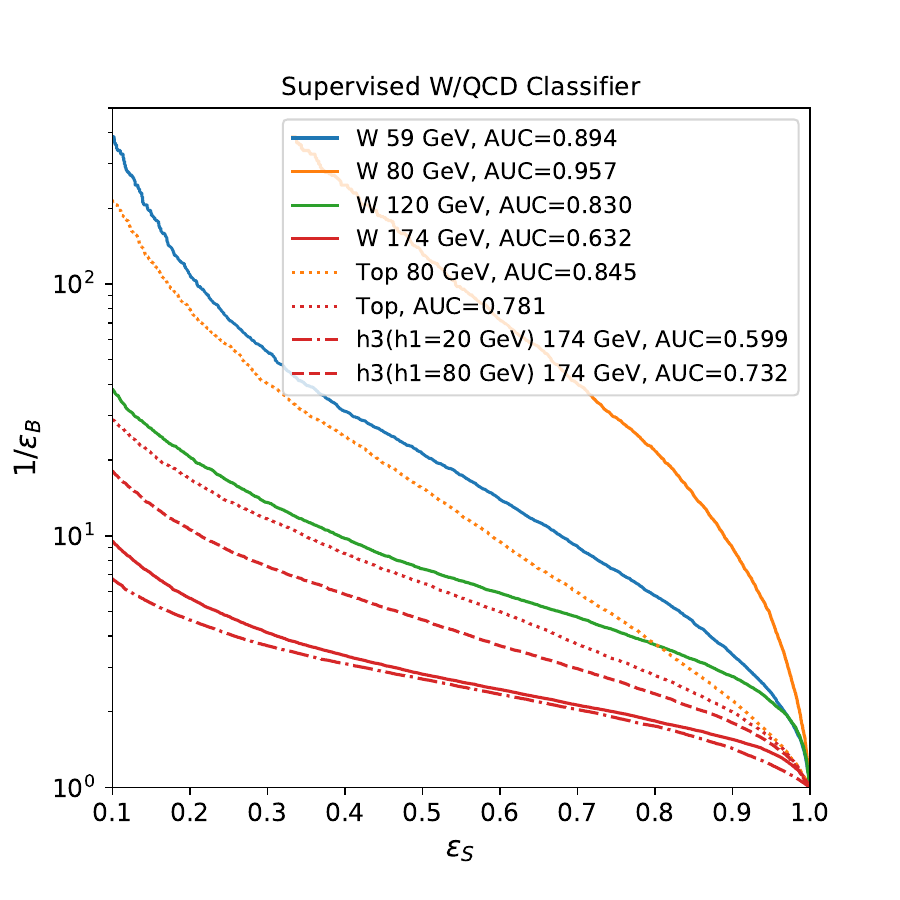}
    \includegraphics[width=0.45\textwidth]{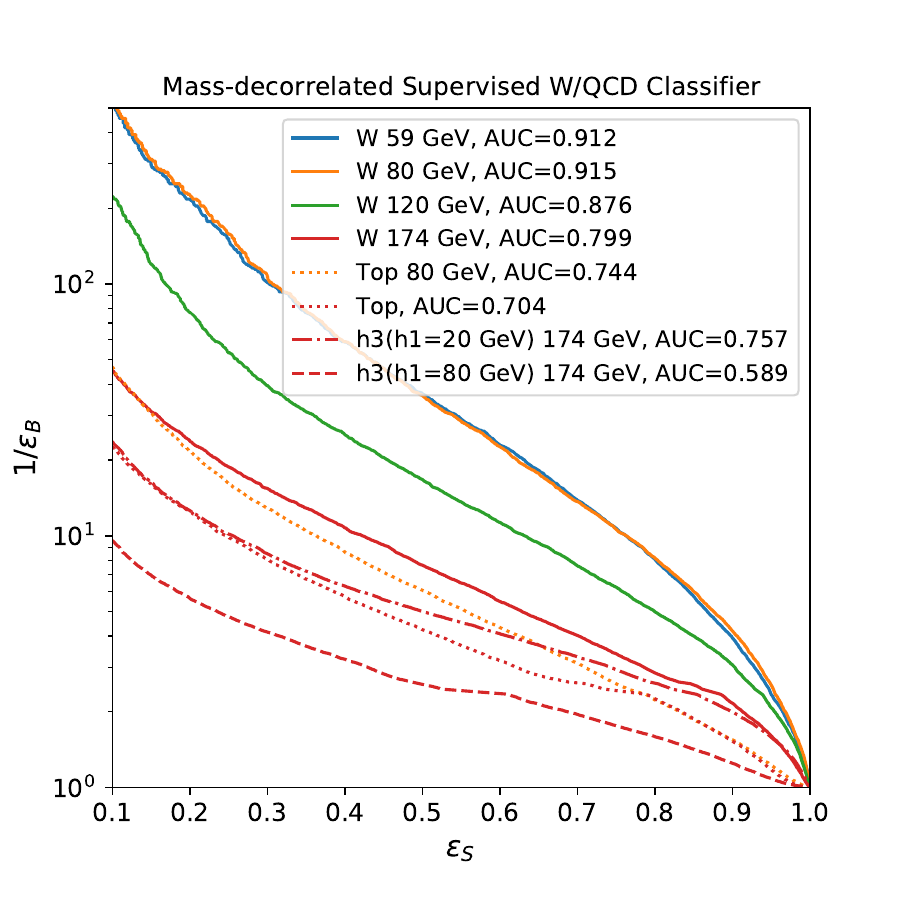}    
    \caption{\textbf{Left}: ROC curves for the supervised $W$/QCD classifier; 
    \textbf{Right}: ROC curves for the mass-decorrelated supervised $W$/QCD classifier.}
    \label{fig:clf_rocs}
\end{figure}

Since we utilized an extra $W$ jet dataset in the semi-supervision approach, it will be interesting to see how the results are compared with supervised $W$/QCD DNN classifiers. We employ a fully-connected DNN architecture (Input(80) $\to$ ReLu(256) $\to$ ReLu(128) $\to$ ReLu(64) $\to$ ReLu(6) $\to$ Sigmoid), and train with the same datasets and input features as in OE-VAE training \footnote{Class weights are employed to balance dataset size of different classes.}. To compare with mass-decorrelated VAE models, we also trained with reweighted samples to decorrelate jet mass for a supervised $W$/QCD tagger. 
%\footnote{We got AUC=0.971 for supervised W/QCD classifier and AUC = 0.947 for mass-decorrelated tagger(full pt range). }.
The mass decorrelation results are shown in Fig. \ref{fig:clf_mdeco}. ROC curves for both taggers are shown in Fig. \ref{fig:clf_rocs}. In general, $W$/QCD classifier tags $W$ jets with different masses quite efficiently. A mass-decorrelated $W$/QCD tagger can also be used to tag other jet types, although with sub-optimal performance. This is due to the remaining transferability of supervised taggers \cite{Cheng:2019isq, Aguilar-Saavedra:2017rzt}. 
%Interestingly, h3(h1=80) has lower performance for supervised taggers, this coincides with our intuition above.

\begin{figure}[htb!]
    \centering
    %\subfigure[]{
    \includegraphics[width=0.3\textwidth]{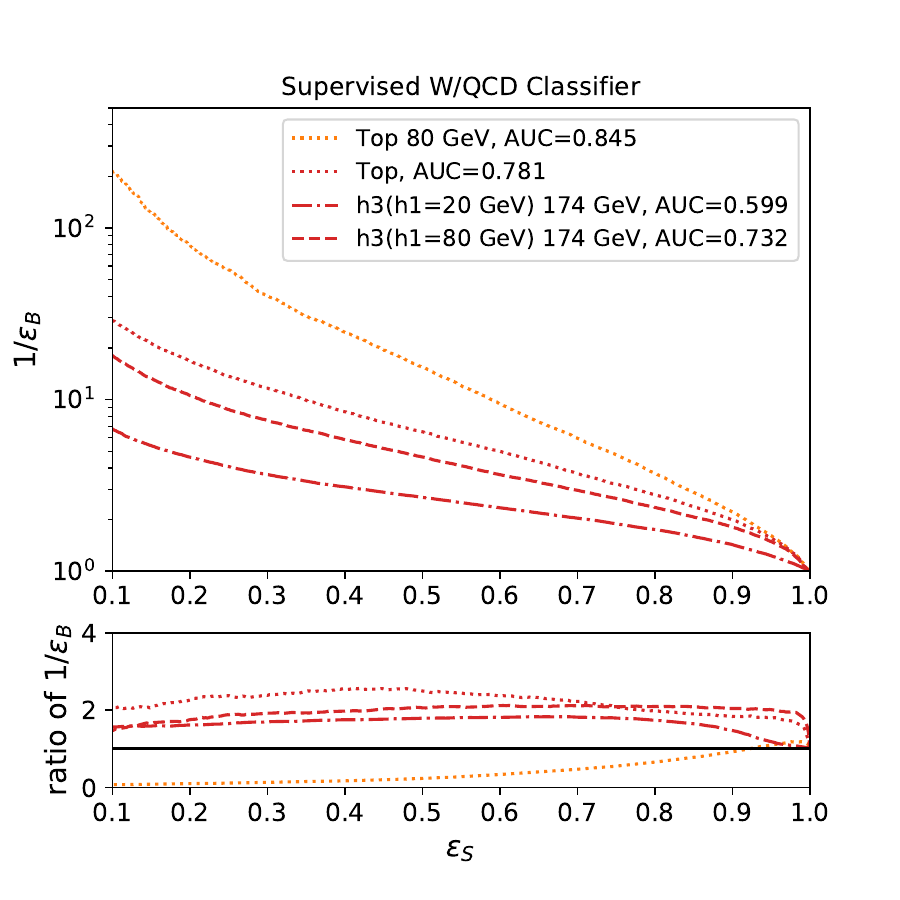}
    %}
    %\subfigure[]{
    \includegraphics[width=0.3\textwidth]{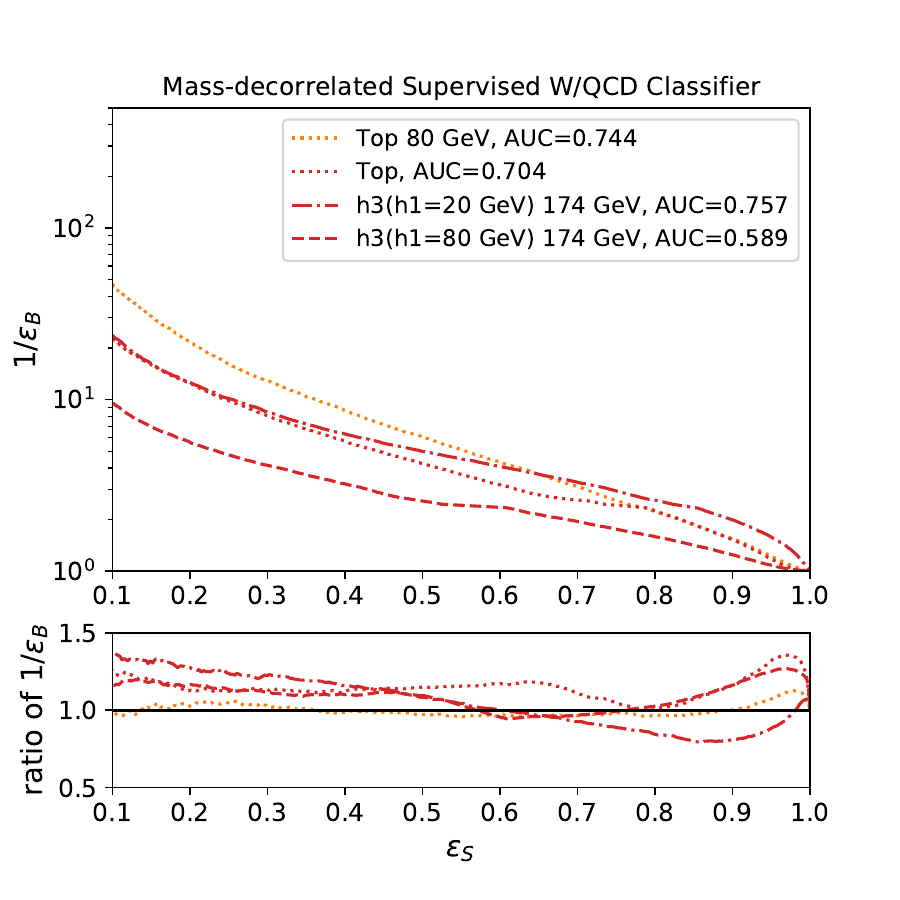}
    %}
    %\subfigure[]{
    \includegraphics[width=0.3\textwidth]{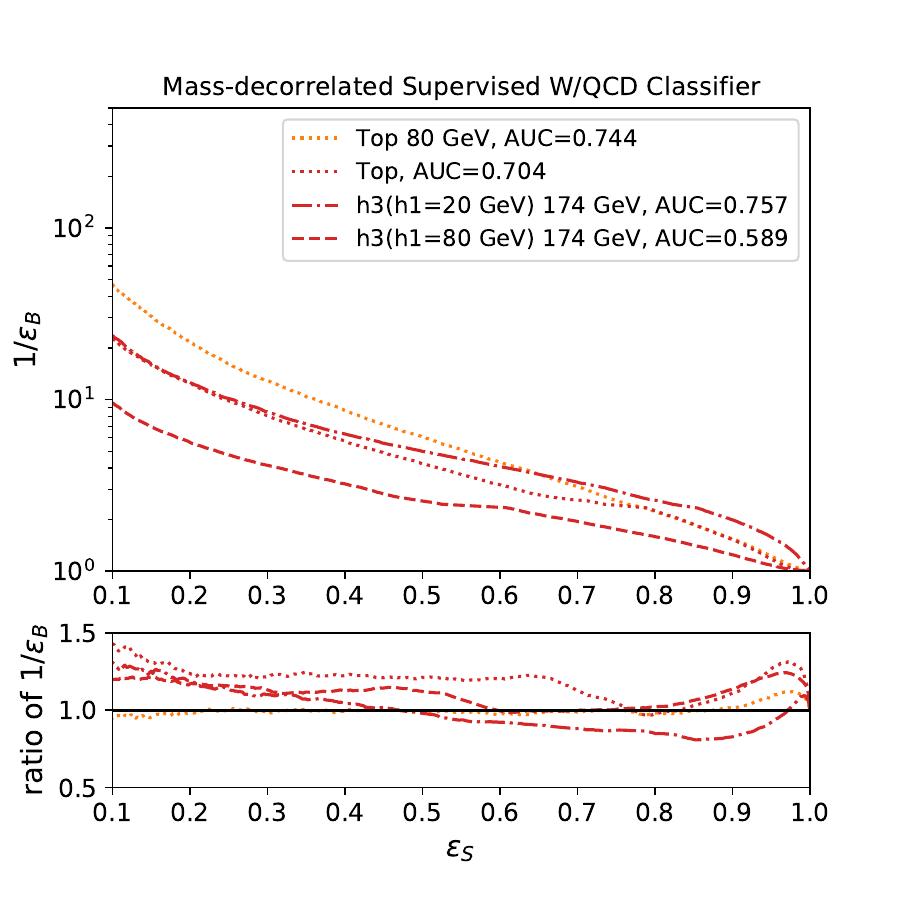}
    %}
    \caption{ROC curves for held-out classes. In the upper part, ROC curves from (mass-decorrelated) $W$/QCD classifier are shown, while in the lower part, the ratios of $1/\epsilon_B$ of (OE-)VAEs to $W$/QCD classifiers are presented.  \textbf{Left}: VAE v.s. $W$/QCD classifier; \textbf{Middle}: MSE-OE v.s. mass-decorrelated $W$/QCD classifier; \textbf{Right}: KL-OE v.s. mass-decorrelated $W$/QCD classifier.}
    \label{fig:rocs_ratios}
\end{figure}

In Fig. \ref{fig:rocs_ratios}, we compare tagging performance on held-out classes (top and Higgs jets) of supervised $W$/QCD classifier and VAE models. On the left panel, we compare the $W$/QCD classifier with the simple VAE for detecting top jets and Higgs jets. VAE generally has better performance regarding these held-out classes, except for the top with a mass of 80 GeV which lies in the sweet spot of a $W$ tagger. However, it's not a completely fair comparison, since both taggers are mass-sculpted. We thus compare mass-decorrelated OE-VAEs with the mass-decorrelated $W$/QCD classifier on the middle and right panels of Fig. \ref{fig:rocs_ratios}. On most of the working points of $\epsilon_S$, the OE-VAEs outperform the $W$/QCD classifier.

\bibliography{ref}
\bibliographystyle{utphys}

\end{document}